\documentclass[fleqn,usenatbib]{mnras}

\usepackage{newtxtext,newtxmath}

\usepackage[T1]{fontenc}

\DeclareRobustCommand{\VAN}[3]{#2}
\let\VANthebibliography\thebibliography
\def\thebibliography{\DeclareRobustCommand{\VAN}[3]{##3}\VANthebibliography}

\usepackage{graphicx}	
\usepackage{amsmath}	
\usepackage{enumitem}
\usepackage{booktabs}

\newcommand{\src}{TCP J1822}

\title[Symbiotic star TCP J18224935-2408280 in outburst]{TCP J18224935-2408280: a symbiotic star identified during outburst}

 \author[Sonith et al.]{
 L. S. Sonith$^{1,2}$\thanks{E-mail: sonith.sls@gmail.com, sonith.ls@iiap.res.in,},
 U. S. Kamath$^{1}$
\\
 $^{1}$Indian Institute of Astrophysics, Koramangala Block II, Bengaluru 560034, Karnataka, India.\\
 $^{2}$Pondicherry University, R.V. Nagar, Kalapet, Puducherry 605014, India.\\
}

\date{Accepted XXX. Received YYY; in original form ZZZ}

\pubyear{2015}

\begin{document}
\label{firstpage}
\pagerange{\pageref{firstpage}--\pageref{lastpage}}
\maketitle

\begin{abstract}

TCP J18224935-2408280 was reported to be in outburst on 2021 May 19. Follow-up spectroscopic observations confirmed that the system was a symbiotic star. We present optical spectra obtained from the Himalayan Chandra Telescope during 2021-22. The early spectra were dominated by Balmer lines, He I lines and high ionization lines such as He II. In the later observations, Raman scattered O VI was also identified. Outburst in the system started as a disc instability, and later the signature of enhanced shell burning and expansion of photospheric radius of the white dwarf was identified. Hence we suggest this outburst is of combination nova type. The post-outburst temperature of the hot component remains above 1.5x10\textsuperscript{5} K  indicating a stable shell burning in the system for a prolonged time after the outburst.
Based on our analysis of archival multiband photometric data, we find that the system contains a cool giant of M1-2 III spectral type with a temperature of $\sim$ 3600K and a radius of $\sim$ 69 R\textsubscript{\(\odot\)}. The pre- and post-outburst light curve shows a periodicity of 631.25 $\pm$ 2.93 d; we consider this as the orbital period. 
\end{abstract}




\section{Introduction}

Symbiotic stars are interacting wide binaries consisting of a cool giant of spectral type M (or K) as a donor star and a hot component, mostly white dwarf (WD) accreting from the giant's wind and surrounded by a circumstellar nebula \citep{2012BaltA..21....5M}. Symbiotic stars manifest a wide variety of variability, from orbital motion to outbursts. Outbursts in symbiotic stars are classified into three, symbiotic novae or slow novae, symbiotic recurrent novae, and classical symbiotic outburst (Z And-type). Symbiotic novae and symbiotic recurrent novae are powered by thermonuclear runaway reactions, whereas classical symbiotic outbursts are believed to be either caused by the release of potential energy from the extra-accreted matter  or due to the increased mass accretion rate followed by the expansion of hot component \citep{2019arXiv190901389M}. Classical symbiotic outbursts are commonly seen feature in symbiotic stars and typically show a 1--3 B mag brightening in the system during the outburst. 

Although classical symbiotic outbursts are one of the most common features of symbiotic stars, we still have limited knowledge about the exact mechanism of these outbursts. In the literature, four different models are proposed to explain these outbursts: 1) Expansion of WD photosphere at a near constant bolometric luminosity due to an increased accretion rate that exceeds steady burning \citep{1976Ap.....12..342T, 1982ApJ...259..244I}, 2) Shell flash or thermal pulse similar to nova and recurrent nova \citep{1983ApJ...273..280K}, 3) Dwarf nova-like outburst due to accretion disc instability \citep{Accretion_disk_model_II_Duschl_1986, Accretion_disk_model_I_Duschl_1986,2002ASPC..261..645M}, 4) Combination nova, where an outburst is initiated by disc instability following an enhanced shell burning \citep{Combination_nova_sokoloski_2006}. In symbiotic stars, it could also be possible to see the same systems showing outbursts with different mechanisms as in AG Peg \citep{AG_Peg_Tomov_2016} or Z And \citep{Combination_nova_sokoloski_2006}. To understand the nature of classical symbiotic outbursts, we require spectroscopic follow-up observations of more systems.

TCP J18224935-2408280 (hereafter referred to as \src) was discovered by Tadashi Kojima, Tsumagoi, Gunma-ken, Japan, on 2021 May 19.683 UT. Discovery was reported in the `Transient Object Followup Reports' pages\footnote{(\url{http://www.cbat.eps.harvard.edu/unconf/followups/J18224935-2408280.html})} of the Central Bureau for Astronomical Telegrams (CBAT). It was suggested to be a symbiotic star outburst by Patrick Schmeer (Saarbrucken-Bischmisheim, Germany) after he found a 2 arcsec nearby Gaia LPV source (Gaia DR2 4089297564356878720) with an approximate orbital period of 800 d. This star is included in the Gaia DR2 catalogue of large-amplitude variables by \cite{2021A&A...648A..44M}. The spectroscopic follow-up observation by \cite{2021ATel14691....1M} on 2021 June 09 showed strong emission lines of H I, He I, O [III], and He II in addition to the K5-M0 continuum. They noted that TCP J1822 is an S-type symbiotic star based on infrared colours; also, the distance and apparent magnitude of the system suggest that it contains a cool component of luminosity class III. The follow-up observations by \cite{2021ATel14692....1A} reached a similar conclusion, and in addition, they have also reported Bowen blend and relatively weak emission lines of Fe II (42, 48, 49 multiplets). Earlier observations by \cite{2021ATel14699....1T} on 2021 June 07, and later observation on June 09 confirm similar nature of \src{} although the identification of the O [III] line is reported weak or absent. We present optical spectroscopic observations of \src{} during 2021--22 and confirm the symbiotic nature of the system and try to understand the nature of the outburst.

\section{Observations} \label{sec:obs}

\subsection{Photometry} \label{subsec:phot}

For understanding the behaviour of \src{} before and during the outburst, we have obtained V and g band photometric data from ASAS-SN sky survey \citep{2014ApJ...788...48S, 2017PASP..129j4502K}, covering the period JD 2457461.83 to JD 2460138.2 (2016 March 14 – 2023 July 12), G, G$_{BP}$ and G$_{RP}$ band magnitudes from Gaia DR3 \citep{2022arXiv220800211G},  covering the period JD 2456913.47 to JD 2457506.75 (2014 September 12 to 2016 April 28). Gaia and ASAS-SN light curves are shown in Fig.~\ref{lc1}.

For obtaining the SED, we have used multiband photometric data available from  Gaia DR3, SkyMapper \citep{2018PASA...35...10W}, TESS Input Catalog (TIC v8.2) \citep{2019AJ....158..138S}, Pan-STARRS1 \citep{2020ApJS..251....7F}, 2MASS \citep{2006AJ....131.1163S}, and WISE \citep{2010AJ....140.1868W}.

\subsection{Spectroscopy}

Low-resolution optical spectra of \src{} were obtained from Himalayan Faint Object Spectrograph Camera (HFOSC) mounted on Himalayan Chandra Telescope (HCT) situated at Indian Astronomical Observatory, Hanle. Observations were carried out between 2021 June 10 and 2021 September 19, using grism 7, having a wavelength range of 3500 to 8000 \AA{} with a resolution of R $\sim$ 1300 and grism 8, having a wavelength range of 5200 to 9000 with a resolution of R $\sim$ 2200. A majority of these observations were carried out in the Target of Opportunity (ToO) mode.
The details of the observations are given in Table \ref{table1}. The data reduction was carried out using a pipeline based on python using pyraf modules, following the standard procedure using different tasks in the Image Reduction and Analysis Facility (IRAF\footnote{IRAF is distributed by the National Optical Astronomy Observatory, which is operated by the Association of Universities for Research in Astronomy (AURA) under a cooperative agreement with the National Science Foundation.}).  Wavelength calibration was carried out for grism 7 and grism 8 using FeAr and FeNe arc lamp spectra, respectively. Feige 110 and Feige 66 were used as standard stars. On the nights in which observation was carried out in the ToO mode, spectrophotometric standards observed on the nearest night were used for correcting the instrumental response.  The response corrected spectra in two grism were scaled to a weighted mean and combined to give the final spectrum. ASAS-SN g-band photometric light curve was used for calibrating the spectra to the absolute flux scale.

\begin{table}
	\caption{Observational log for spectroscopic data obtained for \src{}.}
	\label{table1}
	\resizebox{1\hsize}{!}{\begin{tabular}{cccccc}
			\hline
			\hline
			&  & \textbf{Exposure}  &   \textbf{Wavelength}   \\
			\textbf{Date} &\textbf{JD}  & \textbf{time}  & \textbf{range} \\
			& & \textbf{(s)} & \textbf{(\AA)} \\
			\hline
			2021-06-10  &2459376.36 & 900 + 900         & 3800-9000  \\[0.25ex]
                2021-07-14  &2459410.25 & 1200 + 1200       & 3800-9000  \\[0.25ex]
                2021-09-19  &2459477.11 & 900 + 900         & 3800-9000  \\[0.25ex]
                2022-03-08 \& 09  &2459648.46 & 2400 + 2400 & 3800-9000  \\[0.25ex]
                2022-04-03  &2459673.46 & 2400              & 3800-7700  \\[0.25ex]
                2022-05-14  &2459714.43 & 2400 + 1200       & 3800-9000  \\[0.25ex]
                2022-08-30  &2459822.17& 1800 + 1200        & 3800-9000  \\[0.25ex]
			\hline
	\end{tabular}}
\end{table}

\section{RESULTS AND DISCUSSIONS}

\subsection{Optical light curve and periodicity}

\begin{figure*}
\begin{center}
\includegraphics[width=2\columnwidth]{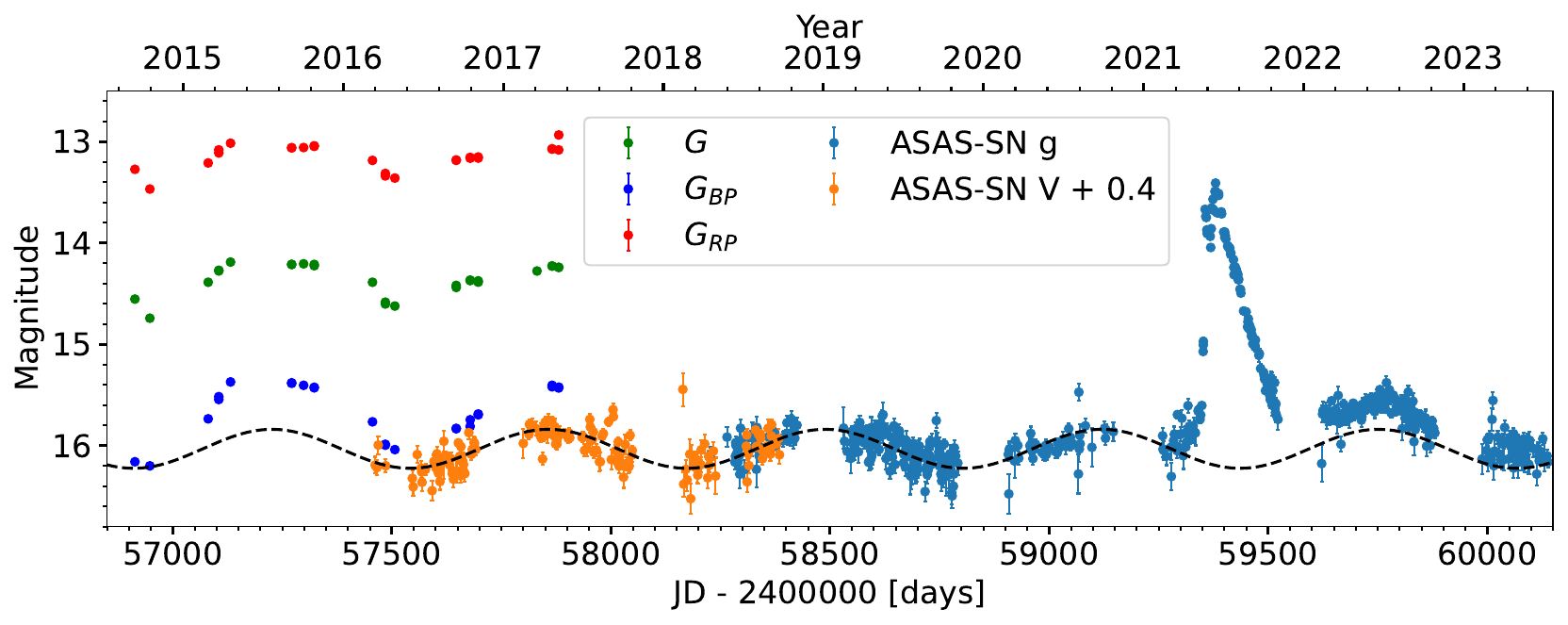}
	\caption{Light curve of \src{} using the Gaia G, G$_{BP}$, G$_{RP}$ magnitudes and ASAS-SN g and V band magnitudes. Periodic behaviour of \src{} is evident in the pre- and post-outburst light curves. The dashed lines show the best fitting sinusoidal curve based on the ephemeris provided in equation~(\ref{eq:ephemeris}).}
	\label{lc1}
\end{center}
\end{figure*}

The ASAS-SN g-band light curve (Fig.~\ref{lc1}) shows that \src{} started brightening on 2021 May 16 and peaked at around 13.5 mag with an increase of 2.2 mag from its quiescent state.  Such a 2--3 mag brightening is often seen in Z-And type symbiotic outbursts. The triangular-shaped outburst peak is similar to the light curve of Z-And outburst of 2000  \citep{Combination_nova_sokoloski_2006}, where it is suggested that a disc instability event causes an initial brightening in the light curve. There followed a decrease of 0.5 mag in the next ten days and a re-brightening to a second maximum, with a broader peak. This follow-up event appears to be related to the nuclear burning on the surface of the WD. After 2021 June 5, the g-band magnitude started to decline again, and \src{} returned to its photometric quiescent state about a year after outburst.

The pre- and post-outburst light curves of \src{} show wave-like variations.  
Using the Lomb-Scargle periodogram (LSP) \citep{1976Ap&SS..39..447L, 1982ApJ...263..835S}, we have obtained periods of 598.95, 618, 598.95 and 609 d corresponding to the highest peaks in the Gaia G, G$_{BP}$, G$_{RP}$ and ASAS-SN V bands, respectively.
A similar analysis using the ASAS-SN g band removing magnitudes during the outburst and shifting the post-outburst quiescent magnitudes to the pre-outburst level gives a period of 629.75 d for the highest peak. 
Furthermore, we estimated the period using multiband data after applying appropriate magnitude shifts and combining them, resulting in the highest LSP peak at 631.69 d, with a false alarm probability of <0.01 per cent.
These periodograms are shown in Fig.~\ref{lsp}. Other peaks obtained from the Gaia data are due to the sampling effect (see appendix \ref{appendix:A}). Additionally, we verified our result for the multiband data using \textit{LombScargleMultiband} function implemented in \textit{astropy} \citep{2018AJ....156..123A}, which gives the same result. Using this period as an initial guess, we have fitted a sinusoidal curve 
and estimated period, light curve minima and associated errors. 
Based on the above analysis, we have obtained an ephemeris of \src{} given by 
\begin{equation}
\label{eq:ephemeris}
JD_{min} = 2457541.69 \pm 5.78 + 631.25 \pm 2.93 \times E
\end{equation}

\begin{figure}
\centering
\includegraphics[width=\columnwidth]{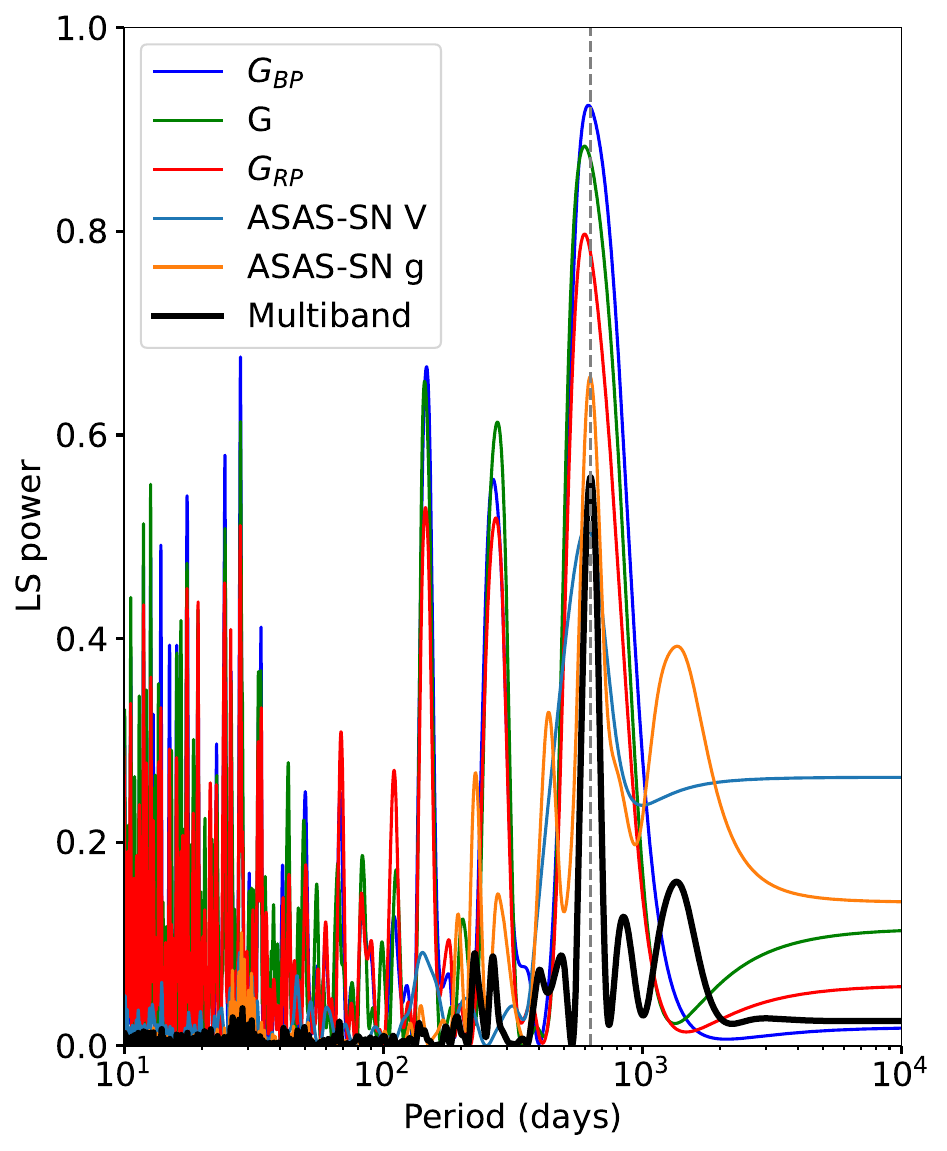}

	\caption{ Comparison of Lomb-Scargle periodograms for \src{} obtained using Gaia and ASAS-SN data, including combined multiband data. LSP give dominant peaks at 598.95, 618, 598.95, 609, 629.75, and 631.69 d in G, G$_{BP}$, G$_{RP}$, ASAS-SN V, ASAS-SN g, and multiband data, respectively. The dashed line indicates a 631.69-day period value. See the text for details.}
	\label{lsp}
\end{figure}

\begin{figure}
\includegraphics[width=\columnwidth]{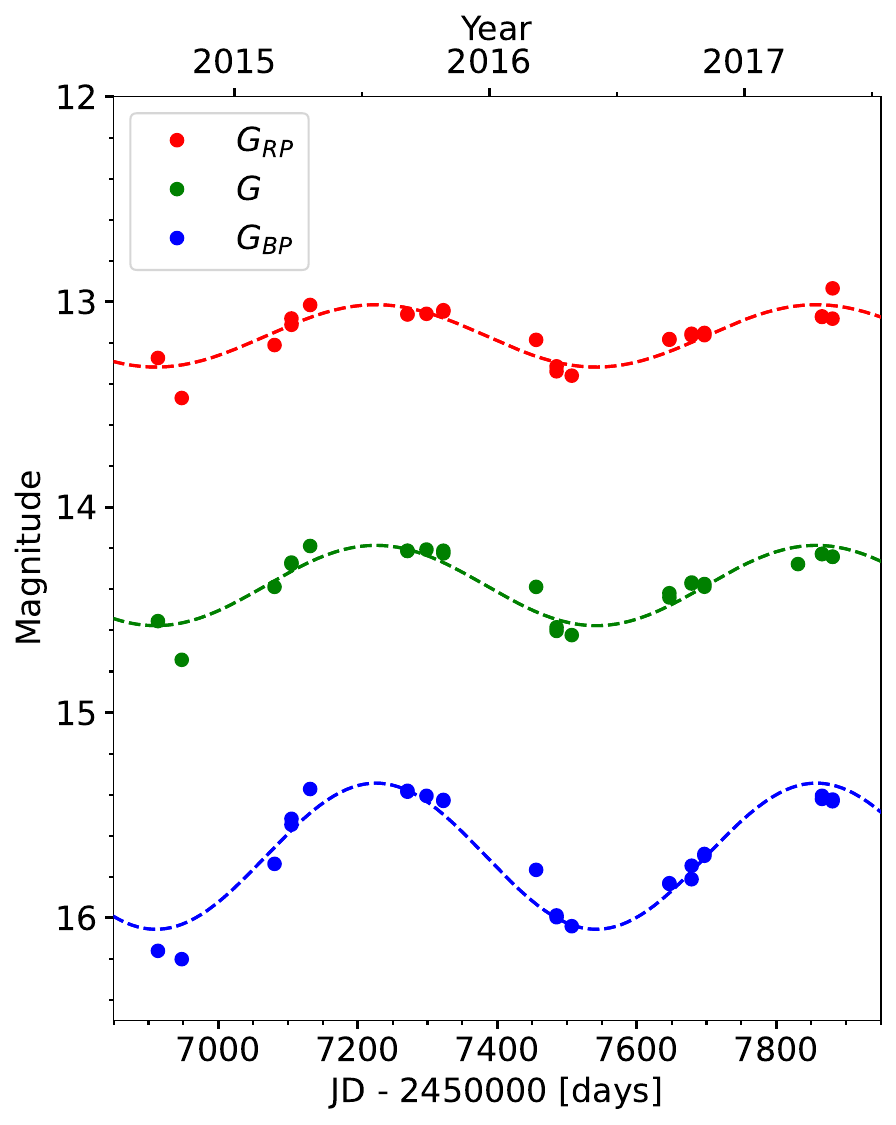}
	\caption{ Gaia G, G$_{BP}$ and G$_{RP}$ band light curves of \src{}. The G, G$_{BP}$ and G$_{RP}$ bands are represented by green, blue and red points, respectively. The dashed lines show the sinusoidal function fitted based on the ephemeris for Gaia G, G$_{BP}$ and G$_{RP}$ bands. See the text for details.}
	\label{lc_fit}
\end{figure}
We attribute this periodicity of 631.25 $\pm$ 2.93 d to the orbital period.
Using the period and phase based on the ephemeris, the  G$_{BP}$, G and G$_{RP}$ data points were fitted using a sinusoidal function by varying the amplitude of the sinusoid  (see Fig.~\ref{lc_fit}). The resultant amplitudes are approximately 0.36, 0.2 and 0.15 mag in the G$_{BP}$, G and G$_{RP}$ bands, respectively. The larger amplitude at the shorter wavelength is indicative of the irradiation of the red giant by the hot WD. This suggests that there was quiescent burning on the suface of the WD in pre-outburst quiescence. Photometry over the next few years will help in refining this value. Future multiband observations would help to delineate the effects of ellipsoidal modulation (which should be prominent in the $I$ band) and irradiation by the hot component ($B$ band).

\subsection{Distance and reddening}
\label{sec_3.3}

From the  Gaia EDR3 parallaxes \citep{2021A&A...649A...1G}, we have estimated the distance to \src{} as $8.1${\raisebox{0.5ex}{\tiny$\substack{+1.6 \\ -1.2}$}} kpc using the \cite{2021AJ....161..147B} method. We have estimated a visual extinction of $A_{\text{v}}$ $\sim$ 1.48 in this direction and for this distance using the 3D map of interstellar dust reddening published by \cite{2019ApJ...887...93G}.  We calculated $A_{\text{v}}$  values using the same procedure for the upper and lower bounds of the distance error, and the results are identical. The  $A_{\text{v}}$ value derived from the \cite{2021AJ....161..147B} approach in the direction of the object does not show significant differences beyond 5 kpc. The map by \cite{2011ApJ...737..103S} indicates that the visual extinction in the direction of TCP J1822 is, $A_{\text{v}} =$ 1.73.

Goodness-of-fit of the astrometric model is -0.16 for Gaia EDR3. A lower than 3 is considered to be a good fit\footnote{(\url{https://gea.esac.esa.int/archive/documentation/GDR2/Gaia_archive/chap_datamodel/sec_dm_main_tables/ssec_dm_gaia_source.html})}. However,
\cite{2021AJ....161..147B} use a probabilistic approach for estimating distances, which relies on priors constructed based on single stars within our Galaxy. It can cause considerable uncertainties for binaries like symbiotic stars.
In this work, we use the distance calculated using EDR3 parallaxes for calculating $A_{\text{v}}$ value and distance prior given for the SED fit (see section \ref{3.3}). Considering that the $A_{\text{v}}$ value is not showing much difference above 5 kpc distance, we are fixing our $A_{\text{v}}$  value to a conservative 1.48 in all the calculations done in this paper. Our spectral type estimation of the cool giant in the system will be hotter by 1-2 spectral sub-types if we take into account a larger $A_{\text{v}}$  value from \cite{2011ApJ...737..103S}. Since our best-fitting spectral type from SED matches well with the quiescent \src{} spectrum above 6000 \AA{}, where the contribution from giant dominates, we find the conservative $A_{\text{v}}$ value we have adopted is suited for subsequent calculations. Reddening corrections were done using the extinction law of \cite{1999PASP..111...63F}.

\subsection{Spectral energy distribution of cool component in \src{}}
\label{3.3}
\begin{figure}
\includegraphics[width=\columnwidth]{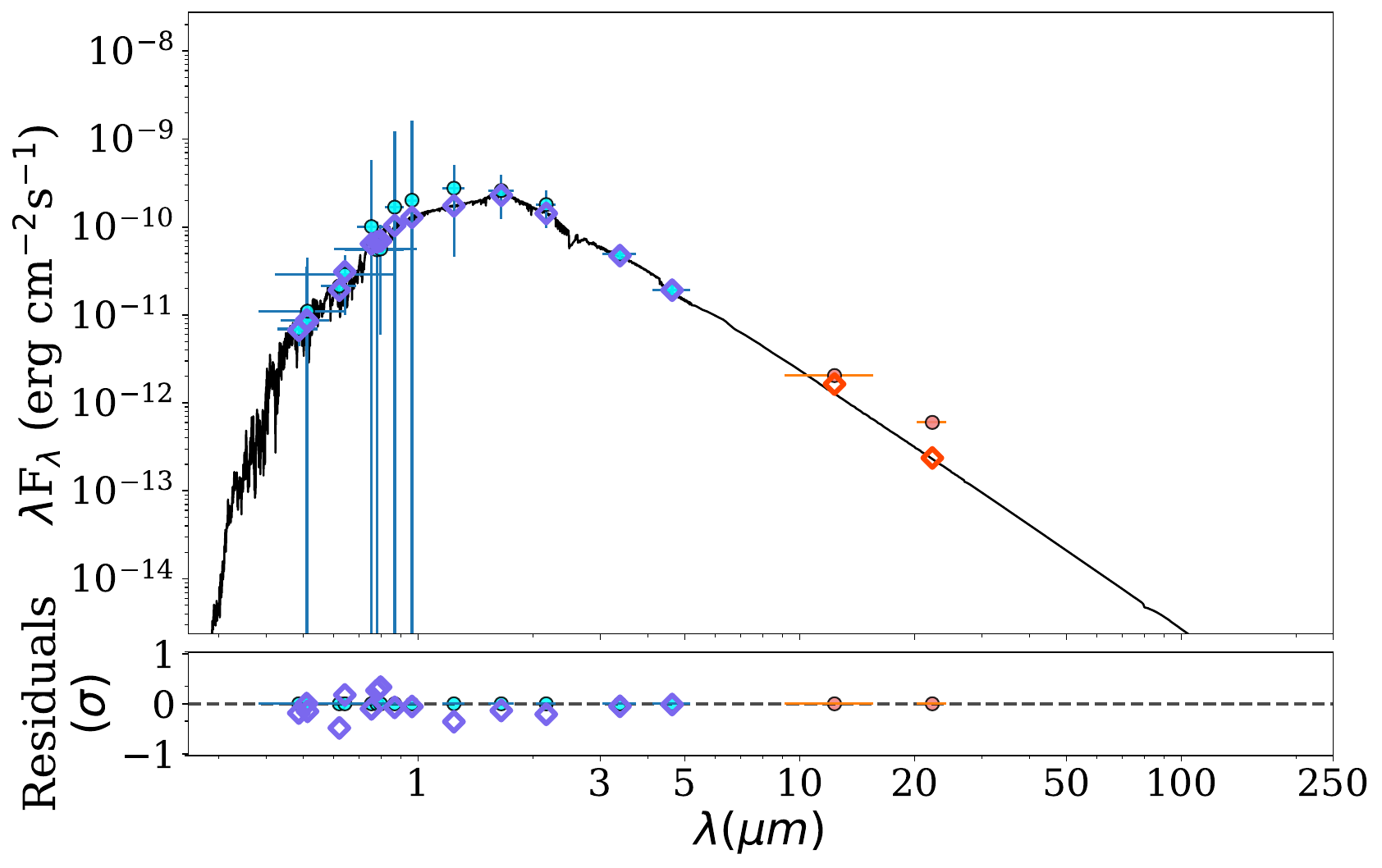}
	\caption{Spectral energy distribution of the \src{} obtained from various bands. Blue points represent the bands used for obtaining the best fit to the SED, as mentioned in section 3.3. WISE W3 and W4 filter magnitudes are over-plotted as orange points. No infrared excess is seen over the stellar continuum. }
	\label{sed}
\end{figure}

\begin{figure}
\includegraphics[width=\columnwidth]{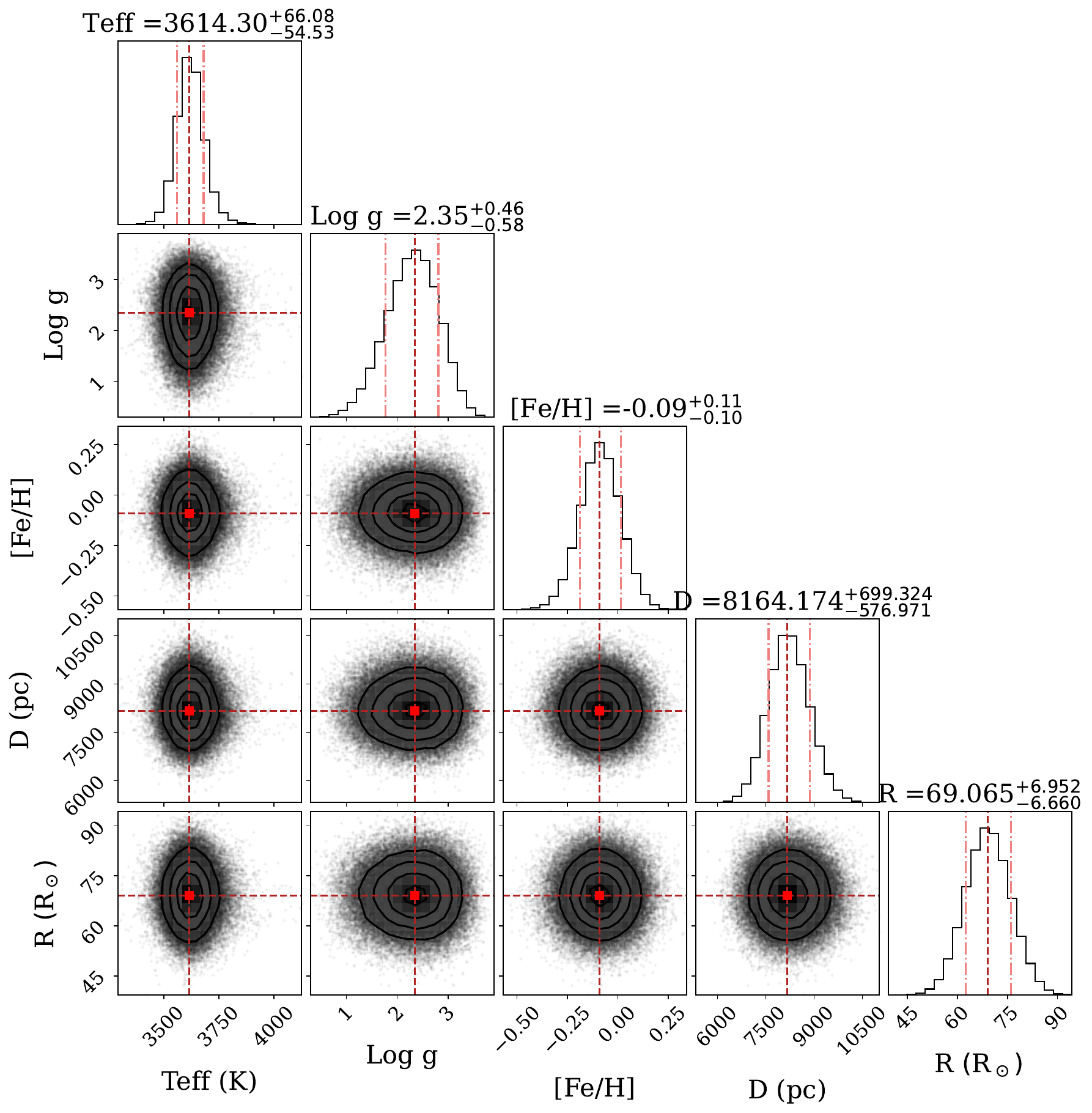}
	\caption{Corner plot indicating different parameters obtained from SED fitting using synthetic model atmospheres. }
	\label{corner_plot}
\end{figure}

The SED of \src{}  obtained using multiband photometric data from Gaia DR3 (G, G$_{BP}$, G$_{RP}$), SkyMapper (g), TESS Input Catalog (TESS), PanSTARRS (g, r, i, z, y), 2MASS (J, H, Ks), and WISE (W1, W2) are shown in Fig.~\ref{sed}.  WISE W3 and W4 bands are shown for representative purposes. We used ARIADNE\footnote{(\url{https://github.com/jvines/astroARIADNE})} \citep{2022MNRAS.513.2719V} to fit the SED using synthetic model atmospheres PHOENIX v2 \citep{PHOENIX_v2}, BT-Settl \citep{Allard_et_al_2012}, BT-NextGen \citep{NextGen_1999, Allard_et_al_2012}, BT-Cond \citep{Allard_et_al_2012}  suits for the temperature range of the cool giant in \src{}. Virtually no excess over the stellar continuum is seen in the WISE W3 and W4 bands, indicating that this is an S-type symbiotic. The fit is quite reasonable, given that the data represent different epochs. For running ARIADNE, we have used the temperature prior based on the Gaia temperature estimate and upper limit. The distance prior is taken from the distance estimate  (see section \ref{sec_3.3}), setting the highest error as the upper limit. The $A_{\text{v}}$ value has been kept fixed at 1.48. We have set a uniform prior for log g from 0 to 4 based on our initial fitting, which results in a radius estimation in the giant star regime. We have used the default prior for radius and metallicity.  The best fit gives a temperature of $3614.30${\raisebox{0.5ex}{\tiny$\substack{+66.08 \\ -54.53}$}}  K, log g = $2.35${\raisebox{0.5ex}{\tiny$\substack{+0.46 \\ -0.58}$}}, radius = $69.06${\raisebox{0.5ex}{\tiny$\substack{+6.95 \\ -6.66}$}}  R\textsubscript{\(\odot\)} and luminosity = $708.2${\raisebox{0.5ex}{\tiny$\substack{+287.6 \\ - 215.7}$}} L\textsubscript{\(\odot\)}, again confirming that the system has a cool giant of M1-2 spectral type. The best fit parameters of SED are shown in Fig.~\ref{corner_plot}. We also estimated mass of the cool component M = $0.9897${\raisebox{0.5ex}{\tiny$\substack{+0.23 \\ -0.07}$}} \(M_\odot\), interpolated from MIST isochrones \citep{mist_1}, with the best-fitting parameters using ARIADNE (see Fig.~\ref{plot:hr} in appendix \ref{appendix:B}).

It should be noted that the temperature, radius and  log g are more accurately constrained, whereas [Fe/H] is indicative, and the obtained SED represents only the mean spectrum of the red giant in \src{}. Since no UV photometry is available, we are unable to probe the nature of the hot component or its orbital variations in a similar manner.

\subsection{Evolution of the optical spectra}
\label{3.4}

\begin{figure*}
\begin{center}
\includegraphics[width=2\columnwidth]{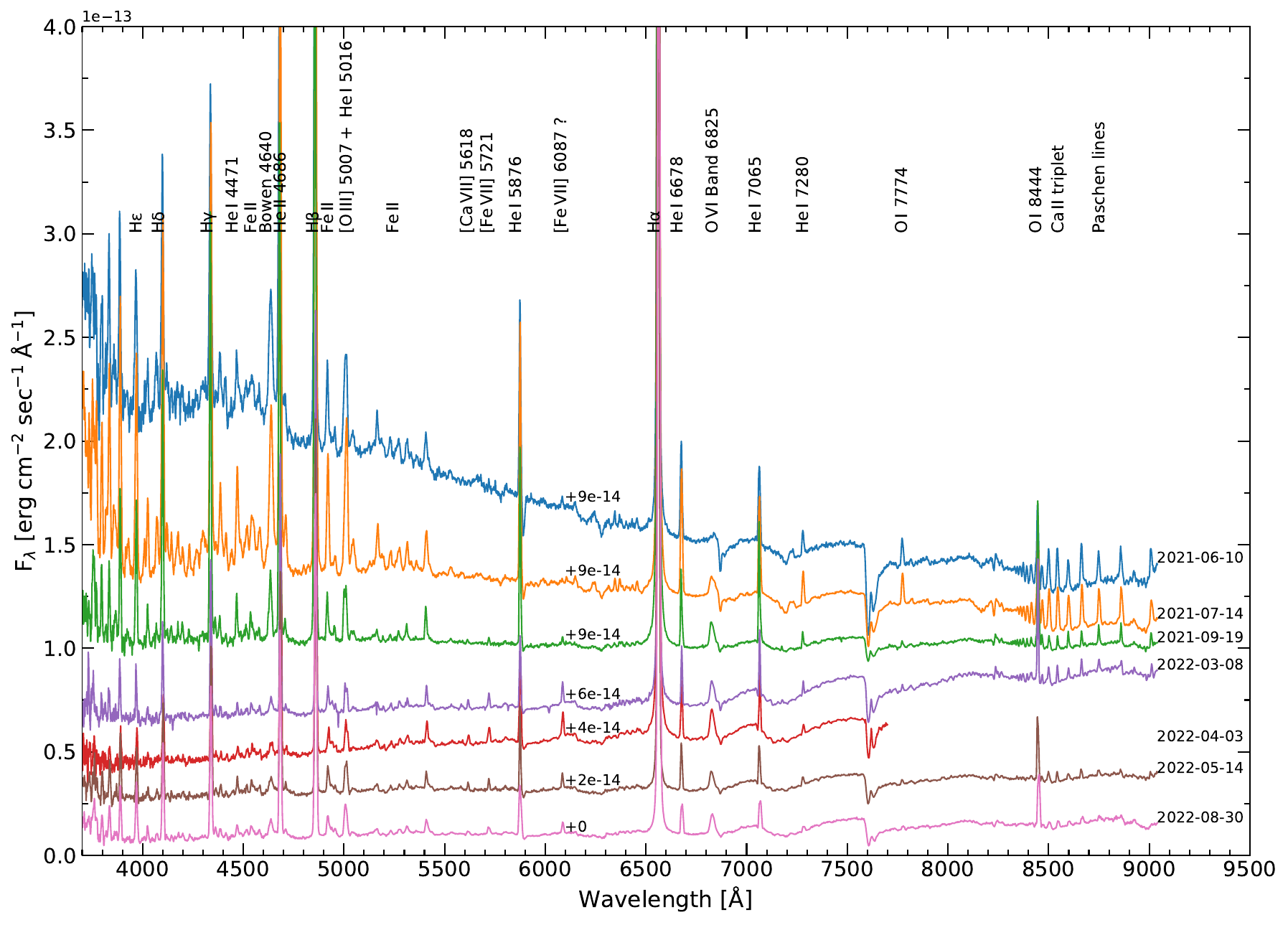}
	\caption{The de-reddened optical low-resolution spectra of symbiotic star \src{} at various epochs of its recent outburst. For clarity, each spectrum has been shifted vertically by the indicated amount.}
	\label{tcp_spectra}
\end{center}
\end{figure*}

\begin{figure}
\centering
\includegraphics[width=\columnwidth]{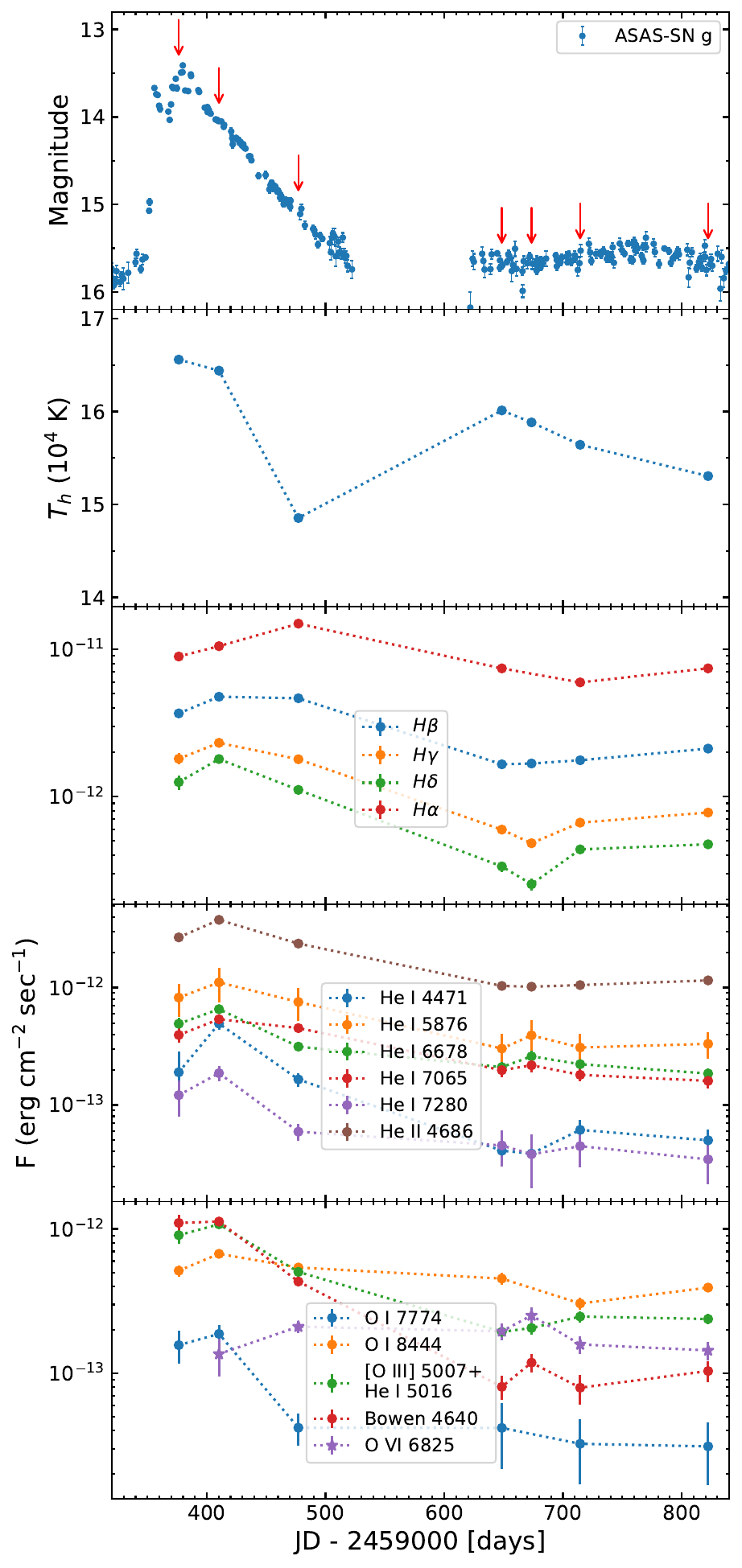}
	\caption{(Top) ASAS-SN g-band light curve of the outburst of \src{}. Epochs of our spectroscopic observations are marked with red arrows. Evolution of the temperature of the hot component and emission line fluxes are shown in the subsequent plots. }
	\label{plot:line}
\end{figure}

We obtained three spectra of \src{} in 2021, during the declining phase of the outburst and four in 2022, after the outburst has almost subsided. Significant changes can be seen in the emission lines and the continuum over this period (see Fig.~\ref{tcp_spectra}). 
The very blue continuum along with strong Balmer and Paschen lines in the spectrum of 2021 June 10 are suggestive of a prominent, hot accretion disc. As the photometry data also indicates, the beginning of the outburst was due to an accretion event. The circumstellar nebula, illuminated by the luminous WD, also contributed to the emission lines.
As the outburst progressed, the blue continuum weakened and the contribution from the red giant in the form of visibility of some TiO bands became increasingly apparent. Typical emission features seen in symbiotic stars such as the Balmer series, He I, O I, Fe II, and high ionization lines such as He II, [O III] and Raman scattered O VI are detected with varying intensity in all the spectra, leaving no doubt about the nature of \src{}. The flux evolution of important lines in the system is shown in Fig.~\ref{plot:line}. H$\alpha$ was saturated for the spectrum obtained on 2022 April 03. The dereddened line fluxes are given in Table \ref{tab:flux}.

The evolution of emission lines of \src{} gives an insight into the nature of the outburst and the hot component present in the system.  All the emission lines fluxes, including Balmer lines, and H I lines, showed an increasing trend in the first (2021 June 10) to the second (2021 July 14) observation and later declined. The O I 7774 \AA{} line is significantly fainter than O I 8444 \AA{}. This may be due to the fluorescence of Ly$\beta$ photons, where Ly$\beta$ photons at 1025.72 \AA{} pump the O I ground state resonance line at 1025.77 \AA{} and later downward cascade to produce 11287\AA{}, 8444\AA{}, and 1304\AA{} lines in emission \citep{Bowen_1947,1995ApJ...439..346K}.

In the quiescent phase spectra (obtained on and after 2022 March 08) all the emission line strengths were reduced, except Raman scattered O VI line and O I 8444 \AA{}. Faint lines of [Ca VII] 5618 \AA{}, [Fe VII] 5721 \AA{}, [Ca V] 6086 \AA{} or [Fe VII] 6087 \AA{} appeared and later reduced their strength in the subsequent observations. High excitation lines emerging in the quiescent spectra indicate that the expanding pseudo-photosphere becomes optically thin (see section \ref{sec:hot_comp}); hence the nebular region is exposed to the heated WD.

\begin{figure}
\centering
\includegraphics[width=\columnwidth]{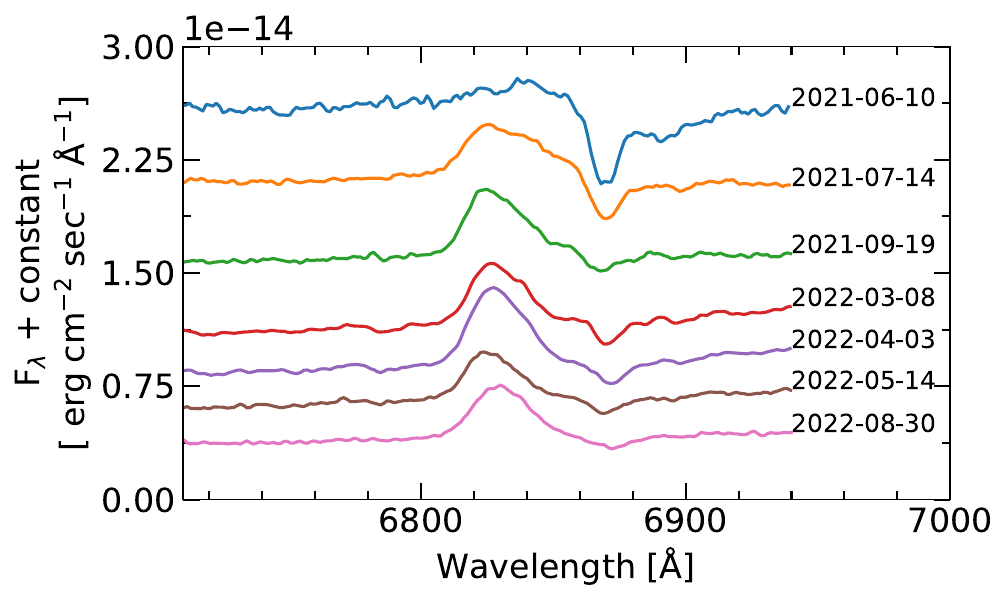}
	\caption{Evolution of Raman scattered O VI band during the \src{} outburst. In the first epoch observation (June 10 2021), g-band light curve was at its peak (see Fig.~\ref{plot:line}) but the Raman scattered O VI band was absent in the optical spectrum. }
	\label{plot:ovi}
\end{figure}

\subsubsection{Raman scattered O VI line}

Raman scattered O VI line is a unique feature that can only be present in a system with neutral hydrogen regions and a hot component which could ionize oxygen to the fifth ionized state \citep{1989A&A...211L..27N}. Hence this feature alone can confirm the symbiotic nature of a system. In our optical spectrum obtained on 2021 June 10 near the g-band maximum, this line was almost invisible. However, it strengthened significantly in the second observation on 2021 July 14 and remained at a similar strength throughout the later observations (see Fig.~\ref{plot:ovi}). Shape of the line was broad on 2021 July 14, but narrowed subsequently. Variations of the Raman scattered O VI line correlated to the optical light curves were also seen in other outbursting symbiotics such as V426 Sge (fig.~7 of \citealp{2020A&A...636A..77S}) and AG Peg (fig.~2 of \citealp{AG_Peg_Tomov_2016}).  However, unlike V426 Sge and AG Peg, in \src{}, after the initial rise, Raman scattered O VI line strength remains almost the same even after the light curve declines. It did not show any decreasing trend while the g-band light curve declined.  The continued line strength for a prolonged period indicates that some extra mass reached the hot component to sustain the shell burning.

\subsubsection{Bowen feature}
The Bowen feature near 4640 \AA{} is seen in all our spectra. This feature was also reported in earlier spectra obtained by  \cite{2021ATel14692....1A, 2021ATel14699....1T}. The strength of the feature reduced as the outburst declined. The Bowen feature is produced when X-rays emitted from a compact source interact with nearby gaseous matter. This indirectly indicates that the system produces X-rays during the outburst --  $\alpha$-type \citep{SySt_in_Xrays_Luna} -- and is consistent with the presence of Raman scattered O VI line in systems showing $\alpha$-type X-ray emission \citep{Akras_survey_syst}. The Bowen feature is also shown by other symbiotic systems like RR Tel and AG Peg \citep{Bowen_excitation_nIII_syst}. These two stars are also reported to show $\alpha$-type and $\beta$-type  X-ray emission, respectively \citep{SySt_in_Xrays_Luna}.  The presence of an enhanced blue continuum during the outburst also indicates the possibility of the $\delta$-type X-ray component in the system, which originates from the inner layer of accretion disc \citep{SySt_in_Xrays_Luna}. Another similar symbiotic with an active accretion disc is MWC 560 (fig.~G1 of \citealp{2020MNRAS.492.3107L}).

\subsection{Nature of the hot component}

\label{sec:hot_comp}

The lower limit of the temperature of the hot component ($T_{\rm h}$) in a symbiotic star can be estimated using the empirical relation ${T}_{\rm h}[1000 {\rm K}]= {\rm \chi}_{\rm max} [{\rm eV}]$ suggested by \cite{1994A&A...282..586M}. This is based on the highest observed ionization potential (${\chi}_{\rm max}$) of an emission line seen in the optical spectrum. Using this, we determine  $T_{\rm h}$ $\gtrsim$ 114 000 K from the presence of Raman scattered O VI band at 6825 \AA{} in the spectra of \src{}, corresponding to the highest ionization potential ${\chi}\textsuperscript{O\textsuperscript{+5}}$  $\sim$ 114 eV.

The hot component is best studied using x-ray and uv observations. In the absence of those, emission lines in the optical are a good proxy for understanding its nature. Considering the hot source to be a blackbody, its temperature and luminosity can be calculated based on H$\beta$, He I and He II lines assuming case B recombination. We have used the relation (\ref{eq:ijima}) derived by \cite{1981ASIC...69..517I}, which is valid for effective temperatures between 70000 to 200000K

\begin{equation}
\label{eq:ijima}
T_{\rm{hot}} (\rm{in\,10^4\,K}) = 19.38 \sqrt{2.22F_{\rm{He \ II \ 4686}} \over4.16F_{\rm{H\,\beta}}+
9.94F_{\rm{He \ I \ 4471}}} + 5.13, 
\end{equation}

The luminosity of the hot component was calculated using equation (8) of \cite{1991AJ....101..637K} and equation (6) given in \cite{1997A&A...327..191M}. Both results match within 25 per cent, and the average value of these estimates is given in Table \ref{tab:hot}. The luminosity estimate using equation (7) of \cite{1997A&A...327..191M}, which is based on H$\beta$ flux, gives a value nearly half of the above. This is not unexpected given that \cite{1997A&A...327..191M} noted these
equations have a factor of $\sim$2 accuracy. A similar effect was reported in the case of Hen 3-860 by \cite{2022MNRAS.510.1404M}, where they have shown H$\beta$ lines having an absorption component seen in high-resolution observations, and hence the flux is getting underestimated. However, we do not see any absorption feature in our low-resolution spectra of \src{}.  

\begin{figure}
\centering
\includegraphics[width=\columnwidth]{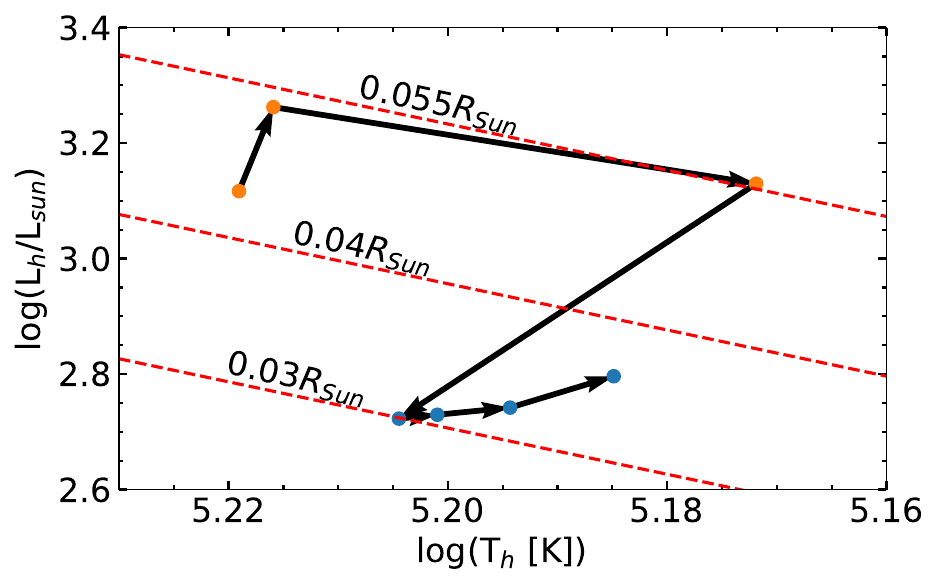}
	\caption{HR diagram showing evolution of the hot component in \src{} from 2021 (orange points) to 2022 (blue points) during the current outburst.}
	\label{plot:hot_hr}
\end{figure}

\begin{table*}
	\centering
	\caption{ The de-reddened absolute fluxes of H$\beta$, He II \,4686\,\AA{}, He I 4471\,\AA{}, He I 5876\ \AA{}, and O VI 6825\,\AA\ line, together with the estimated luminosity, temperature
and radius of the TCP hot component.}
	\label{tab:hot}
	\begin{tabular}{ccccccccccc} 
		\hline		
\multicolumn{1}{c}{Date}&\multicolumn{1}{c}{JD}&\multicolumn{
5}{c}{Flux \,ergs\,cm$^\mathrm{-2}$\,s$^\mathrm{-1}$}& 
\multicolumn{1}{c}{$T_\mathrm{ h}$}&
\multicolumn{1}{c}{$L_\mathrm{h}$}&
\multicolumn{1}{c}{$R_\mathrm{h}$}\\ 

&\multicolumn{1}{c}{since 2400000}& \multicolumn{1}{c}{He II\,4686\,\AA} & 
\multicolumn{1}{c}{H$\beta$} & \multicolumn{1}{c}{He I 4471\,\AA} & \multicolumn{1}{c}{He I 5876\,\AA} &
\multicolumn{1}{c}{O VI 6825\,\AA} &\multicolumn{1}{c}{[10$^3$\,K]} & \multicolumn{1}{c}{[$L_{\sun}$]} & 
\multicolumn{1}{c}{[$R_{\sun}$]} \\
		\hline

2021-06-10      & 59376.36& 2.69e-12&  3.67e-12 &  1.9e-13 &   8.24e-13 & --          & 165.6&    1310.0& 0.044\\ 
2021-07-14      & 59410.25& 3.79e-12&  4.76e-12 & 4.94e-13 &   1.11e-12 & 1.36e-13    & 164.4&    1830.0& 0.053\\ 
2021-09-19      & 59477.11& 2.38e-12&  4.65e-12 & 1.65e-13 &   7.56e-13 & 2.11e-13    & 148.6&    1350.0& 0.055\\ 
2022-03-08 \& 09& 59648.46& 1.04e-12&  1.65e-12 &  4.1e-14 &   3.02e-13 & 1.95e-13    & 160.1&     530.0& 0.03\\ 
2022-04-03      & 59673.45& 1.02e-12&  1.68e-12 &  3.8e-14 &   3.92e-13 & 2.52e-13    & 158.8&     540.0& 0.031\\ 
2022-05-14      & 59714.43& 1.05e-12&  1.76e-12 &  6.1e-14 &   3.09e-13 & 1.59e-13    & 156.4&     550.0& 0.032\\ 
2022-08-30      & 59822.17& 1.15e-12&  2.12e-12 &    5e-14 &   3.32e-13 & 1.44e-13    & 153.1&     630.0& 0.036\\
		\hline
	\end{tabular}
\end{table*}

\begin{figure}
\centering
\includegraphics[width=0.5\textwidth]{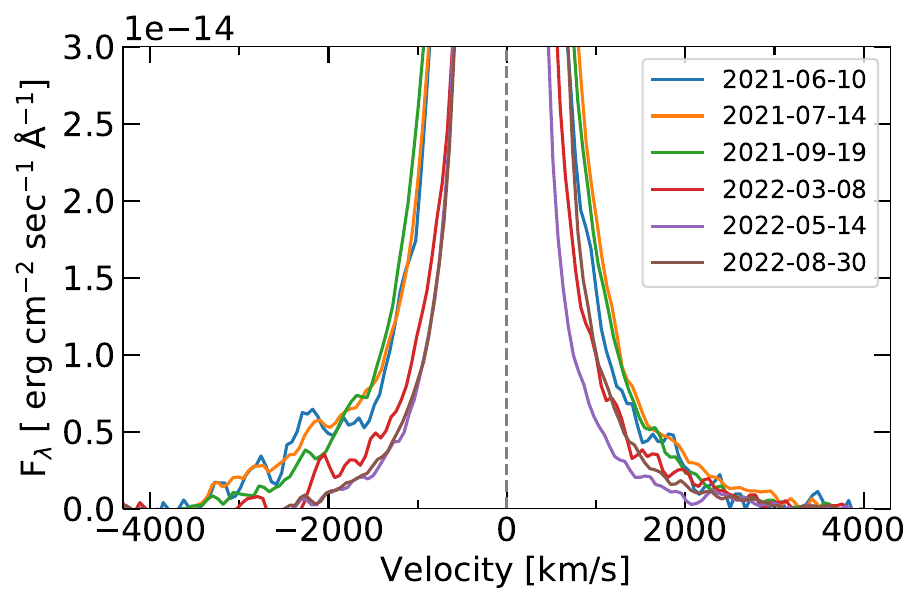}
	\caption{ Broadening of H$\alpha$ line during outburst of \src{}. The H$\alpha$ line is plotted after subtracting the local continuum. Enhanced bluer wing of H$\alpha$ line is an indication of outflow while the outburst happened in the system. }
	\label{plot:halpha}
\end{figure}

The blackbody assumption also allows determination of the radius, which is given in Table \ref{tab:hot}. From Fig.~\ref{plot:hot_hr} it is evident that radius of the hot component showed an increasing trend during the outburst decline. There is an enhancement in the blue wings of H$\alpha$ early during the outburst; the line width is also broader (see Fig.~\ref{plot:halpha}). The radius suddenly dropped when \src{} reached the quiescence phase (last four observations). The increase in radius was due to the physical expansion of the photosphere caused by excess burning on the surface of the WD. As the photosphere expanded, the temperature dropped. During the quiescence phase, the expanded shell became optically thin, and hence radius showed a sudden drop, which means we started seeing closer to the WD again.

H$\alpha$ wing profiles presented in Fig.~\ref{plot:halpha}  are obtained by subtracting the local continuum using \textit{ fit\_continuum} function in \textit{Specutils} \citep{specutils_1.10.0}. We see H$\alpha$ wings as broad as $\sim$3500 km/s in the blue region and $\sim$3000 km/s in the red region for the first three spectra taken during the outburst. The H$\alpha$ wings in the blue region are stronger than those in the red region. Line broadening has been reported in past outbursts in AG Peg (fig.~3 in \citealp{AG_Peg_Tomov_2016}) and V426 Sge (fig.~3 in \citealp{2020A&A...636A..77S}).
However, in the case of AG Peg and V426 Sge, velocities of H$\alpha$ wing profiles are lower ($\leq$1500 km/s) compared to what we observe in TCP J1822. These broadenings are due to an increased outflow during the outburst.

\subsection{Nature of the outburst}

The optical eruption observed in \src{} shows an amplitude of around 2.5 mag in the ASAS-SN g-band and is similar to Z And-type outburst seen in classical symbiotic stars. They show a brightening of 1-3 mag with time scales from months to years (e.g. Z And, CI Cyg, and AG Dra). In addition, spectroscopic observations of \src{} after the optical maximum show that forbidden lines (e.g. [O III], [Fe VII]; see section \ref{3.4}) are weak and lower density lines (e.g. [O II], [S II], [N II]) are absent which favours the classical symbiotic outburst interpretation. (see e.g. CI Cyg - \citealp{1991AJ....101..637K}; AG Dra - \citealp{1995AJ....109.1289M}; LIN 9 - \citealp{2014MNRAS.444L..11M}).
The multi-peak light curve of \src{}, with a sharp rise during the outburst, resembles that of Z And, which showed a combination nova outburst in 2000 \citep{Combination_nova_sokoloski_2006}.
Dominance of the blue continuum, strong Balmer and Paschen lines in the early outburst, and the nature of the light curve indicate that some sort of disc instability was responsible. This probably deposited additional matter on the already burning WD, causing the second peak in the light curve (see Fig.~\ref{plot:line}). Dwarf nova-like disc instability as a triggering mechanism for Z And outburst is also examined in the theoretical model by \cite{disk_instabilities_2018MNRAS.481.5422B}. It is estimated that a high accretion rate of the order of 10\textsuperscript{-6} M\textsubscript{\(\odot\)} yr\textsuperscript{-1} is required for such a scenario to be feasible in a symbiotic star like Z And. \cite{disk_instabilities_2018MNRAS.481.5422B} suggest that such an enhancement in mass transfer could be attributed to the magnetic activity on the surface of the giant as suggested by \cite{2008MNRAS.385..445L}. In this scenario, the increased mass transfer could act as a trigger mechanism for enhanced shell burning.
The continuum observed during the outburst of \src{} is derived from multiple components, including the nebula, accretion disc, and WD. Understanding the individual contributions of each component requires rigorous modelling, which is beyond the scope of this paper.

The presence of high ionization lines like He II 4686 and Raman-scattered 6825 from outburst through near-quiescence indicates that the WD continued to burn matter on its surface. This could give rise to detectable soft X-rays. X-ray data would be needed to understand the relative contributions of steady nuclear burning and accretion in the system. 
The strength of the Raman scattered O VI line remains high even after nearly a year after the outburst declined, indicating that enough material reached the surface of the hot component to maintain the shell burning for a prolonged time. From the ASAS-SN g band light curve (Fig.~\ref{lc1}), it is seen that the post-outburst magnitude is brighter than the pre-outburst magnitudes, which further confirms our finding.

After returning to quiescence, \src{} exhibits a temperature of above 10\textsuperscript{5} K, luminosity of order 10\textsuperscript{3} L\textsubscript{\(\odot\)}, which is typical for the hot component in quiescently burning symbiotic stars (fig.~4 in \citealp{2003ASPC..303....9M}, and \citealp{2019arXiv190901389M}).

\section{Conclusions}


\begin{enumerate}[leftmargin=*,rightmargin=0ex,itemsep=0.8ex,label=(\roman*).]
\item The optical spectrum of \src{} shows Balmer series lines, O I, He I, and high excitation lines such as He II, O[III], Raman scattered O VI and TiO band heads from the cool component which unambiguously confirm the symbiotic nature of the system.
\item We probed the nature of the cool component in the system using multiband SED and found that the system contains an M1-2 III spectral-type star having a temperature of $\sim$ 3600K, radius of $\sim$ 69 R\textsubscript{\(\odot\)} and luminosity of $\sim$ 700 L\textsubscript{\(\odot\)}.    
\item \src{} shows a combination nova type outburst where the outburst begins as accretion disc instability during the first peak of the light curve and then enhances the shell burning in the system, which is correlated with the radius increase of WD photosphere.  
\item The pre- and post-outburst light curve of \src{} shows a 631.25 $\pm$ 2.93 day periodic variation, which most probably originates from the orbital motion of the system. 
\item The post-outburst temperature of the hot component remains above  1.5x10\textsuperscript{5} K, indicating a stable shell burning in the system for a prolonged time after the outburst. The strength of Raman scattered O VI band and elevated post-outburst ASAS-SN g band
magnitude compared to pre-outburst also confirms the same. These findings collectively suggest an enhanced mass transfer during the outburst. 

\end{enumerate}

\section*{Acknowledgements}

We thank the anonymous reviewer for thoroughly reading our paper and providing insightful comments and suggestions. We thank the staff of IAO, Hanle and Centre For Research \& Education in Science \& Technology (CREST), Hosakote, that made these observations possible. The facilities at IAO and CREST are operated by the Indian Institute of Astrophysics, Bangalore. We thank all the observers of HCT for accommodating some time for Target of Opportunity (ToO) observations. We also thank the HCT time allocation committee (HTAC) for the time and support during ToO and regular observations. This research made use of Astropy,\footnote{http://www.astropy.org} a community-developed core Python package for Astronomy \citep{2018AJ....156..123A}.

\section*{Data Availability}

ASAS-SN and Gaia photometric light curves are available from the ASAS-SN Archive\footnote{https://asas-sn.osu.edu/} and Gaia Archive\footnote{https://gea.esac.esa.int/archive/}, respectively. Spectroscopic data will be provided by the corresponding author upon reasonable request. 
 



\bibliographystyle{mnras}
\bibliography{tcpj1822} 

\begin{thebibliography}{}
\makeatletter
\relax
\def\mn@urlcharsother{\let\do\@makeother \do\$\do\&\do\#\do\^\do\_\do\%\do\~}
\def\mn@doi{\begingroup\mn@urlcharsother \@ifnextchar [ {\mn@doi@} {\mn@doi@[]}}
\def\mn@doi@[#1]#2{\def\@tempa{#1}\ifx\@tempa\@empty \href {http://dx.doi.org/#2} {doi:#2}\else \href {http://dx.doi.org/#2} {#1}\fi \endgroup}
\def\mn@eprint#1#2{\mn@eprint@#1:#2::\@nil}
\def\mn@eprint@arXiv#1{\href {http://arxiv.org/abs/#1} {{\tt arXiv:#1}}}
\def\mn@eprint@dblp#1{\href {http://dblp.uni-trier.de/rec/bibtex/#1.xml} {dblp:#1}}
\def\mn@eprint@#1:#2:#3:#4\@nil{\def\@tempa {#1}\def\@tempb {#2}\def\@tempc {#3}\ifx \@tempc \@empty \let \@tempc \@tempb \let \@tempb \@tempa \fi \ifx \@tempb \@empty \def\@tempb {arXiv}\fi \@ifundefined {mn@eprint@\@tempb}{\@tempb:\@tempc}{\expandafter \expandafter \csname mn@eprint@\@tempb\endcsname \expandafter{\@tempc}}}

\bibitem[\protect\citeauthoryear{{Akras}, {Guzman-Ramirez}, {Leal-Ferreira}  \& {Ramos-Larios}}{{Akras} et~al.}{2019}]{Akras_survey_syst}
{Akras} S.,  {Guzman-Ramirez} L.,  {Leal-Ferreira} M.~L.,   {Ramos-Larios} G.,  2019, \mn@doi [\apjs] {10.3847/1538-4365/aaf88c}, \href {https://ui.adsabs.harvard.edu/abs/2019ApJS..240...21A} {240, 21}

\bibitem[\protect\citeauthoryear{{Allard}, {Homeier}  \& {Freytag}}{{Allard} et~al.}{2012}]{Allard_et_al_2012}
{Allard} F.,  {Homeier} D.,   {Freytag} B.,  2012, \mn@doi [Philosophical Transactions of the Royal Society of London Series A] {10.1098/rsta.2011.0269}, \href {https://ui.adsabs.harvard.edu/abs/2012RSPTA.370.2765A} {370, 2765}

\bibitem[\protect\citeauthoryear{{Astropy Collaboration} et~al.,}{{Astropy Collaboration} et~al.}{2018}]{2018AJ....156..123A}
{Astropy Collaboration} et~al., 2018, \mn@doi [\aj] {10.3847/1538-3881/aabc4f}, \href {https://ui.adsabs.harvard.edu/abs/2018AJ....156..123A} {156, 123}

\bibitem[\protect\citeauthoryear{{Aydi}, {Sokolovsky}, {Strader}, {Chomiuk}  \& {Kawash}}{{Aydi} et~al.}{2021}]{2021ATel14692....1A}
{Aydi} E.,  {Sokolovsky} K.~V.,  {Strader} J.,  {Chomiuk} L.,   {Kawash} A.,  2021, The Astronomer's Telegram, \href {https://ui.adsabs.harvard.edu/abs/2021ATel14692....1A} {14692, 1}

\bibitem[\protect\citeauthoryear{{Bailer-Jones}, {Rybizki}, {Fouesneau}, {Demleitner}  \& {Andrae}}{{Bailer-Jones} et~al.}{2021}]{2021AJ....161..147B}
{Bailer-Jones} C.~A.~L.,  {Rybizki} J.,  {Fouesneau} M.,  {Demleitner} M.,   {Andrae} R.,  2021, \mn@doi [\aj] {10.3847/1538-3881/abd806}, \href {https://ui.adsabs.harvard.edu/abs/2021AJ....161..147B} {161, 147}

\bibitem[\protect\citeauthoryear{{Bollimpalli}, {Hameury}  \& {Lasota}}{{Bollimpalli} et~al.}{2018}]{disk_instabilities_2018MNRAS.481.5422B}
{Bollimpalli} D.~A.,  {Hameury} J.~M.,   {Lasota} J.~P.,  2018, \mn@doi [\mnras] {10.1093/mnras/sty2555}, \href {https://ui.adsabs.harvard.edu/abs/2018MNRAS.481.5422B} {481, 5422}

\bibitem[\protect\citeauthoryear{{Bowen}}{{Bowen}}{1947}]{Bowen_1947}
{Bowen} I.~S.,  1947, \mn@doi [\pasp] {10.1086/125951}, \href {https://ui.adsabs.harvard.edu/abs/1947PASP...59..196B} {59, 196}

\bibitem[\protect\citeauthoryear{{Dotter}}{{Dotter}}{2016}]{mist_1}
{Dotter} A.,  2016, \mn@doi [\apjs] {10.3847/0067-0049/222/1/8}, \href {https://ui.adsabs.harvard.edu/abs/2016ApJS..222....8D} {222, 8}

\bibitem[\protect\citeauthoryear{{Duschl}}{{Duschl}}{1986a}]{Accretion_disk_model_II_Duschl_1986}
{Duschl} W.~J.,  1986a, \aap, \href {https://ui.adsabs.harvard.edu/abs/1986A&A...163...61D} {163, 61}

\bibitem[\protect\citeauthoryear{{Duschl}}{{Duschl}}{1986b}]{Accretion_disk_model_I_Duschl_1986}
{Duschl} W.~J.,  1986b, \aap, \href {https://ui.adsabs.harvard.edu/abs/1986A&A...163...56D} {163, 56}

\bibitem[\protect\citeauthoryear{{Earl} et~al.,}{{Earl} et~al.}{2023}]{specutils_1.10.0}
{Earl} N.,  et~al., 2023, {astropy/specutils: v1.10.0}, Zenodo, \mn@doi{10.5281/zenodo.7803739}

\bibitem[\protect\citeauthoryear{{Eriksson}, {Johansson}, {Wahlgren}, {Veenhuizen}, {Munari}  \& {Siviero}}{{Eriksson} et~al.}{2005}]{Bowen_excitation_nIII_syst}
{Eriksson} M.,  {Johansson} S.,  {Wahlgren} G.~M.,  {Veenhuizen} H.,  {Munari} U.,   {Siviero} A.,  2005, \mn@doi [\aap] {10.1051/0004-6361:20042174}, \href {https://ui.adsabs.harvard.edu/abs/2005A&A...434..397E} {434, 397}

\bibitem[\protect\citeauthoryear{{Fitzpatrick}}{{Fitzpatrick}}{1999}]{1999PASP..111...63F}
{Fitzpatrick} E.~L.,  1999, \mn@doi [\pasp] {10.1086/316293}, \href {https://ui.adsabs.harvard.edu/abs/1999PASP..111...63F} {111, 63}

\bibitem[\protect\citeauthoryear{{Flewelling} et~al.,}{{Flewelling} et~al.}{2020}]{2020ApJS..251....7F}
{Flewelling} H.~A.,  et~al., 2020, \mn@doi [\apjs] {10.3847/1538-4365/abb82d}, 251, 7

\bibitem[\protect\citeauthoryear{{Gaia Collaboration} et~al.,}{{Gaia Collaboration} et~al.}{2021}]{2021A&A...649A...1G}
{Gaia Collaboration} et~al., 2021, \mn@doi [\aap] {10.1051/0004-6361/202039657}, \href {https://ui.adsabs.harvard.edu/abs/2021A&A...649A...1G} {649, A1}

\bibitem[\protect\citeauthoryear{{Gaia Collaboration} et~al.,}{{Gaia Collaboration} et~al.}{2022}]{2022arXiv220800211G}
{Gaia Collaboration} et~al., 2022, arXiv e-prints, \href {https://ui.adsabs.harvard.edu/abs/2022arXiv220800211G} {p. arXiv:2208.00211}

\bibitem[\protect\citeauthoryear{{Green}, {Schlafly}, {Zucker}, {Speagle}  \& {Finkbeiner}}{{Green} et~al.}{2019}]{2019ApJ...887...93G}
{Green} G.~M.,  {Schlafly} E.,  {Zucker} C.,  {Speagle} J.~S.,   {Finkbeiner} D.,  2019, \mn@doi [\apj] {10.3847/1538-4357/ab5362}, \href {https://ui.adsabs.harvard.edu/abs/2019ApJ...887...93G} {887, 93}

\bibitem[\protect\citeauthoryear{{Hauschildt}, {Allard}  \& {Baron}}{{Hauschildt} et~al.}{1999}]{NextGen_1999}
{Hauschildt} P.~H.,  {Allard} F.,   {Baron} E.,  1999, \mn@doi [\apj] {10.1086/306745}, \href {https://ui.adsabs.harvard.edu/abs/1999ApJ...512..377H} {512, 377}

\bibitem[\protect\citeauthoryear{{Husser}, {Wende-von Berg}, {Dreizler}, {Homeier}, {Reiners}, {Barman}  \& {Hauschildt}}{{Husser} et~al.}{2013}]{PHOENIX_v2}
{Husser} T.~O.,  {Wende-von Berg} S.,  {Dreizler} S.,  {Homeier} D.,  {Reiners} A.,  {Barman} T.,   {Hauschildt} P.~H.,  2013, \mn@doi [\aap] {10.1051/0004-6361/201219058}, \href {https://ui.adsabs.harvard.edu/abs/2013A&A...553A...6H} {553, A6}

\bibitem[\protect\citeauthoryear{{Iben}}{{Iben}}{1982}]{1982ApJ...259..244I}
{Iben} I. J.,  1982, \mn@doi [\apj] {10.1086/160164}, \href {https://ui.adsabs.harvard.edu/abs/1982ApJ...259..244I} {259, 244}

\bibitem[\protect\citeauthoryear{Iijima}{Iijima}{1981}]{1981ASIC...69..517I}
Iijima T.,  1981, in Photometric and Spectroscopic Binary Systems. p.~517

\bibitem[\protect\citeauthoryear{{Kastner} \& {Bhatia}}{{Kastner} \& {Bhatia}}{1995}]{1995ApJ...439..346K}
{Kastner} S.~O.,  {Bhatia} A.~K.,  1995, \mn@doi [\apj] {10.1086/175178}, \href {https://ui.adsabs.harvard.edu/abs/1995ApJ...439..346K} {439, 346}

\bibitem[\protect\citeauthoryear{{Kenyon} \& {Truran}}{{Kenyon} \& {Truran}}{1983}]{1983ApJ...273..280K}
{Kenyon} S.~J.,  {Truran} J.~W.,  1983, \mn@doi [\apj] {10.1086/161367}, \href {https://ui.adsabs.harvard.edu/abs/1983ApJ...273..280K} {273, 280}

\bibitem[\protect\citeauthoryear{{Kenyon}, {Oliversen}, {Mikolajewska}, {Mikolajewski}, {Stencel}, {Garcia}  \& {Anderson}}{{Kenyon} et~al.}{1991}]{1991AJ....101..637K}
{Kenyon} S.~J.,  {Oliversen} N.~A.,  {Mikolajewska} J.,  {Mikolajewski} M.,  {Stencel} R.~E.,  {Garcia} M.~R.,   {Anderson} C.~M.,  1991, \mn@doi [\aj] {10.1086/115712}, \href {https://ui.adsabs.harvard.edu/abs/1991AJ....101..637K} {101, 637}

\bibitem[\protect\citeauthoryear{{Kochanek} et~al.,}{{Kochanek} et~al.}{2017}]{2017PASP..129j4502K}
{Kochanek} C.~S.,  et~al., 2017, \mn@doi [\pasp] {10.1088/1538-3873/aa80d9}, \href {https://ui.adsabs.harvard.edu/abs/2017PASP..129j4502K} {129, 104502}

\bibitem[\protect\citeauthoryear{{Leibowitz} \& {Formiggini}}{{Leibowitz} \& {Formiggini}}{2008}]{2008MNRAS.385..445L}
{Leibowitz} E.~M.,  {Formiggini} L.,  2008, \mn@doi [\mnras] {10.1111/j.1365-2966.2008.12847.x}, \href {https://ui.adsabs.harvard.edu/abs/2008MNRAS.385..445L} {385, 445}

\bibitem[\protect\citeauthoryear{{Lomb}}{{Lomb}}{1976}]{1976Ap&SS..39..447L}
{Lomb} N.~R.,  1976, \mn@doi [\apss] {10.1007/BF00648343}, \href {https://ui.adsabs.harvard.edu/abs/1976Ap&SS..39..447L} {39, 447}

\bibitem[\protect\citeauthoryear{{Lucy} et~al.,}{{Lucy} et~al.}{2020}]{2020MNRAS.492.3107L}
{Lucy} A.~B.,  et~al., 2020, \mn@doi [\mnras] {10.1093/mnras/stz3595}, \href {https://ui.adsabs.harvard.edu/abs/2020MNRAS.492.3107L} {492, 3107}

\bibitem[\protect\citeauthoryear{{Luna}, {Sokoloski}, {Mukai}  \& {Nelson}}{{Luna} et~al.}{2013}]{SySt_in_Xrays_Luna}
{Luna} G.~J.~M.,  {Sokoloski} J.~L.,  {Mukai} K.,   {Nelson} T.,  2013, \mn@doi [\aap] {10.1051/0004-6361/201220792}, \href {https://ui.adsabs.harvard.edu/abs/2013A&A...559A...6L} {559, A6}

\bibitem[\protect\citeauthoryear{{Merc}, {Galis}, {Charbonnel}, {Garde}, {Le Du}, {Mulato}  \& {Petit}}{{Merc} et~al.}{2021}]{2021ATel14691....1M}
{Merc} J.,  {Galis} R.,  {Charbonnel} S.,  {Garde} O.,  {Le Du} P.,  {Mulato} L.,   {Petit} T.,  2021, The Astronomer's Telegram, \href {https://ui.adsabs.harvard.edu/abs/2021ATel14691....1M} {14691, 1}

\bibitem[\protect\citeauthoryear{{Merc}, {G{\'a}lis}, {Wolf}, {Velez}, {Bohlsen}  \& {Barlow}}{{Merc} et~al.}{2022}]{2022MNRAS.510.1404M}
{Merc} J.,  {G{\'a}lis} R.,  {Wolf} M.,  {Velez} P.,  {Bohlsen} T.,   {Barlow} B.~N.,  2022, \mn@doi [\mnras] {10.1093/mnras/stab3512}, \href {https://ui.adsabs.harvard.edu/abs/2022MNRAS.510.1404M} {510, 1404}

\bibitem[\protect\citeauthoryear{{Miko{\l}ajewska}}{{Miko{\l}ajewska}}{2003}]{2003ASPC..303....9M}
{Miko{\l}ajewska} J.,  2003, in {Corradi} R.~L.~M.,  {Mikolajewska} J.,   {Mahoney} T.~J.,  eds,  Astronomical Society of the Pacific Conference Series Vol. 303, Symbiotic Stars Probing Stellar Evolution. p.~9 (\mn@eprint {arXiv} {astro-ph/0210489}), \mn@doi{10.48550/arXiv.astro-ph/0210489}

\bibitem[\protect\citeauthoryear{{Miko{\l}ajewska}}{{Miko{\l}ajewska}}{2012}]{2012BaltA..21....5M}
{Miko{\l}ajewska} J.,  2012, \mn@doi [Baltic Astronomy] {10.1515/astro-2017-0352}, \href {https://ui.adsabs.harvard.edu/abs/2012BaltA..21....5M} {21, 5}

\bibitem[\protect\citeauthoryear{{Mikolajewska}, {Kenyon}, {Mikolajewski}, {Garcia}  \& {Polidan}}{{Mikolajewska} et~al.}{1995}]{1995AJ....109.1289M}
{Mikolajewska} J.,  {Kenyon} S.~J.,  {Mikolajewski} M.,  {Garcia} M.~R.,   {Polidan} R.~S.,  1995, \mn@doi [\aj] {10.1086/117361}, \href {https://ui.adsabs.harvard.edu/abs/1995AJ....109.1289M} {109, 1289}

\bibitem[\protect\citeauthoryear{{Mikolajewska}, {Acker}  \& {Stenholm}}{{Mikolajewska} et~al.}{1997}]{1997A&A...327..191M}
{Mikolajewska} J.,  {Acker} A.,   {Stenholm} B.,  1997, \aap, \href {https://ui.adsabs.harvard.edu/abs/1997A&A...327..191M} {327, 191}

\bibitem[\protect\citeauthoryear{{Mikolajewska}, {Kolotilov}, {Shenavrin}  \& {Yudin}}{{Mikolajewska} et~al.}{2002}]{2002ASPC..261..645M}
{Mikolajewska} J.,  {Kolotilov} E.~A.,  {Shenavrin} V.~I.,   {Yudin} B.~F.,  2002, in {G{\"a}nsicke} B.~T.,  {Beuermann} K.,   {Reinsch} K.,  eds,  Astronomical Society of the Pacific Conference Series Vol. 261, The Physics of Cataclysmic Variables and Related Objects. p.~645

\bibitem[\protect\citeauthoryear{{Miszalski}, {Mikolajewska}  \& {Udalski}}{{Miszalski} et~al.}{2014}]{2014MNRAS.444L..11M}
{Miszalski} B.,  {Mikolajewska} J.,   {Udalski} A.,  2014, \mn@doi [\mnras] {10.1093/mnrasl/slu098}, \href {https://ui.adsabs.harvard.edu/abs/2014MNRAS.444L..11M} {444, L11}

\bibitem[\protect\citeauthoryear{{Mowlavi} et~al.,}{{Mowlavi} et~al.}{2021}]{2021A&A...648A..44M}
{Mowlavi} N.,  et~al., 2021, \mn@doi [\aap] {10.1051/0004-6361/202039450}, \href {https://ui.adsabs.harvard.edu/abs/2021A&A...648A..44M} {648, A44}

\bibitem[\protect\citeauthoryear{{Munari}}{{Munari}}{2019}]{2019arXiv190901389M}
{Munari} U.,  2019, arXiv e-prints, \href {https://ui.adsabs.harvard.edu/abs/2019arXiv190901389M} {p. arXiv:1909.01389}

\bibitem[\protect\citeauthoryear{{Murset} \& {Nussbaumer}}{{Murset} \& {Nussbaumer}}{1994}]{1994A&A...282..586M}
{Murset} U.,  {Nussbaumer} H.,  1994, \aap, \href {https://ui.adsabs.harvard.edu/abs/1994A&A...282..586M} {282, 586}

\bibitem[\protect\citeauthoryear{{Nussbaumer}, {Schmid}  \& {Vogel}}{{Nussbaumer} et~al.}{1989}]{1989A&A...211L..27N}
{Nussbaumer} H.,  {Schmid} H.~M.,   {Vogel} M.,  1989, \aap, \href {https://ui.adsabs.harvard.edu/abs/1989A&A...211L..27N} {211, L27}

\bibitem[\protect\citeauthoryear{{Scargle}}{{Scargle}}{1982}]{1982ApJ...263..835S}
{Scargle} J.~D.,  1982, \mn@doi [\apj] {10.1086/160554}, \href {https://ui.adsabs.harvard.edu/abs/1982ApJ...263..835S} {263, 835}

\bibitem[\protect\citeauthoryear{{Schlafly} \& {Finkbeiner}}{{Schlafly} \& {Finkbeiner}}{2011}]{2011ApJ...737..103S}
{Schlafly} E.~F.,  {Finkbeiner} D.~P.,  2011, \mn@doi [\apj] {10.1088/0004-637X/737/2/103}, \href {https://ui.adsabs.harvard.edu/abs/2011ApJ...737..103S} {737, 103}

\bibitem[\protect\citeauthoryear{{Shappee} et~al.,}{{Shappee} et~al.}{2014}]{2014ApJ...788...48S}
{Shappee} B.~J.,  et~al., 2014, \mn@doi [\apj] {10.1088/0004-637X/788/1/48}, \href {https://ui.adsabs.harvard.edu/abs/2014ApJ...788...48S} {788, 48}

\bibitem[\protect\citeauthoryear{{Skopal} et~al.,}{{Skopal} et~al.}{2020}]{2020A&A...636A..77S}
{Skopal} A.,  et~al., 2020, \mn@doi [\aap] {10.1051/0004-6361/201937199}, \href {https://ui.adsabs.harvard.edu/abs/2020A&A...636A..77S} {636, A77}

\bibitem[\protect\citeauthoryear{{Skrutskie} et~al.,}{{Skrutskie} et~al.}{2006}]{2006AJ....131.1163S}
{Skrutskie} M.~F.,  et~al., 2006, \mn@doi [\aj] {10.1086/498708}, \href {https://ui.adsabs.harvard.edu/abs/2006AJ....131.1163S} {131, 1163}

\bibitem[\protect\citeauthoryear{{Sokoloski} et~al.,}{{Sokoloski} et~al.}{2006}]{Combination_nova_sokoloski_2006}
{Sokoloski} J.~L.,  et~al., 2006, \mn@doi [\apj] {10.1086/498206}, \href {https://ui.adsabs.harvard.edu/abs/2006ApJ...636.1002S} {636, 1002}

\bibitem[\protect\citeauthoryear{{Stassun} et~al.,}{{Stassun} et~al.}{2019}]{2019AJ....158..138S}
{Stassun} K.~G.,  et~al., 2019, \mn@doi [\aj] {10.3847/1538-3881/ab3467}, 158, 138

\bibitem[\protect\citeauthoryear{{Taguchi}, {Maehara}, {Fujii}  \& {Kato}}{{Taguchi} et~al.}{2021}]{2021ATel14699....1T}
{Taguchi} K.,  {Maehara} H.,  {Fujii} M.,   {Kato} T.,  2021, The Astronomer's Telegram, \href {https://ui.adsabs.harvard.edu/abs/2021ATel14699....1T} {14699, 1}

\bibitem[\protect\citeauthoryear{{Tomov}, {Stoyanov}  \& {Zamanov}}{{Tomov} et~al.}{2016}]{AG_Peg_Tomov_2016}
{Tomov} T.~V.,  {Stoyanov} K.~A.,   {Zamanov} R.~K.,  2016, \mn@doi [\mnras] {10.1093/mnras/stw2012}, \href {https://ui.adsabs.harvard.edu/abs/2016MNRAS.462.4435T} {462, 4435}

\bibitem[\protect\citeauthoryear{{Tutukov} \& {Yungel'Son}}{{Tutukov} \& {Yungel'Son}}{1976}]{1976Ap.....12..342T}
{Tutukov} A.~V.,  {Yungel'Son} L.~R.,  1976, \mn@doi [Astrophysics] {10.1007/BF01003331}, \href {https://ui.adsabs.harvard.edu/abs/1976Ap.....12..342T} {12, 342}

\bibitem[\protect\citeauthoryear{{Vines} \& {Jenkins}}{{Vines} \& {Jenkins}}{2022}]{2022MNRAS.513.2719V}
{Vines} J.~I.,  {Jenkins} J.~S.,  2022, \mn@doi [\mnras] {10.1093/mnras/stac956}, \href {https://ui.adsabs.harvard.edu/abs/2022MNRAS.513.2719V} {513, 2719}

\bibitem[\protect\citeauthoryear{{Wolf} et~al.,}{{Wolf} et~al.}{2018}]{2018PASA...35...10W}
{Wolf} C.,  et~al., 2018, \mn@doi [\pasa] {10.1017/pasa.2018.5}, \href {https://ui.adsabs.harvard.edu/abs/2018PASA...35...10W} {35, e010}

\bibitem[\protect\citeauthoryear{{Wright} et~al.,}{{Wright} et~al.}{2010}]{2010AJ....140.1868W}
{Wright} E.~L.,  et~al., 2010, \mn@doi [\aj] {10.1088/0004-6256/140/6/1868}, 140, 1868

\makeatother
\end{thebibliography}




\appendix



\section{Comparison of LSP from simulated and observed Gaia data}
\label{appendix:A}

\begin{figure}
\centering
\includegraphics[width=0.5\textwidth]{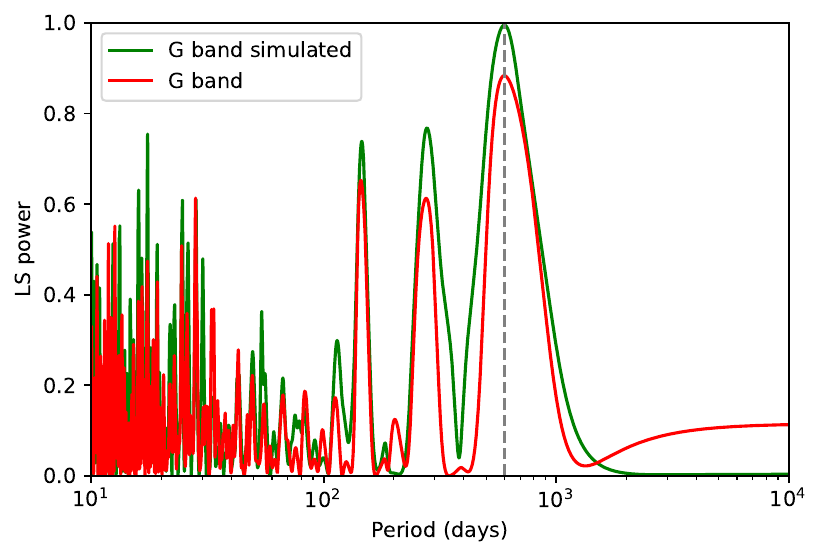}
	\caption{Lomb-scargle periodogram generated using simulated Gaia G magnitudes and observed Gaia G magnitudes are shown in the figure. Simulated data points are created on the same observed epochs to check the sampling effect. We assumed a sinusoidal variation in the G band light curve and used the same period we obtained from the observed G magnitudes, 598.95 d.}
	\label{plot:tcp_lsp_sim}
\end{figure}

The Gaia data points we obtain show groupings with a 140-day interval. Within each group, the two or three observations have a 24 d separation. This sampling shows up as aliases seen in the LSP as minor peaks around 24, 140, and 270 d.   To confirm this, we have simulated a data set with 598.95 d period with similar sampling as Gaia data. We were able to reproduce the peaks we were getting around the above periods, which are shown in Fig.~\ref{plot:tcp_lsp_sim}.

\section{MIST isochrones fit using ARIADNE}
\label{appendix:B}

\begin{figure}
\centering
\includegraphics[width=0.5\textwidth]{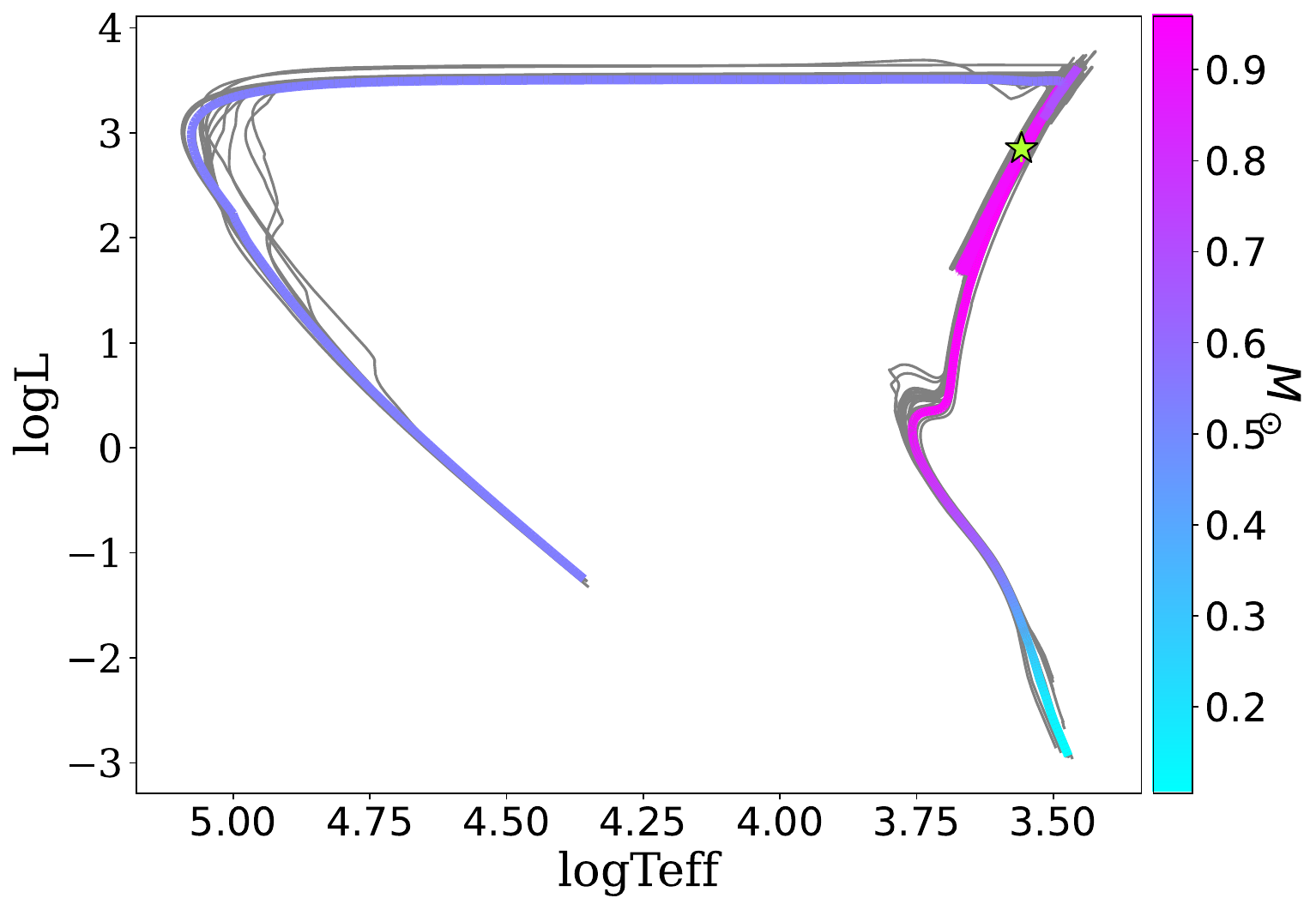}
	\caption{HR diagram of cool component in \src{} over plotted with MIST isochrones.}
	\label{plot:hr}
\end{figure}

ARIADNE also provides mass of the star by interpolating MIST isochrones, using the best-fitting parameters obtained from SED as input (see Fig.~\ref{plot:hr}).

\section{Error propagation in line flux}
\label{appendix:C}
\begin{enumerate}[leftmargin=*,rightmargin=0ex,itemsep=0.8ex,label=(\roman*).]
\item Initially, the spectra, extracted using IRAF, are accompanied by error estimates.
\item Subsequently, this spectrum is calibrated into an absolute flux scale, utilizing photometric zero points from the ASAS-SN g band, with the photometry-derived flux error propagated to spectra. After this step, the continuum has $\sim$ 10 \%  error.
\item The process of estimating line flux involves subtracting the local continuum, which introduces additional error into the continuum subtracted spectrum.
\item During the process of fitting the line, the statistical error resulting from the fitting procedure is combined with the error associated with the continuum-subtracted line and propagated to line flux measurement.
\item Strong lines demonstrate errors under 10\%, whereas moderately strong lines encompass errors spanning 10\% to 30\%. Some weak lines and those that got weaker during spectral evolution have shown errors reaching around 50 \%. In the case of the He I 5876 line, it has a higher uncertainty ($\sim$30\%) due to blending with nearby Na doublet lines; the flux estimate was derived subsequent to deblending.

\end{enumerate}

\begin{table*}
    \centering
    \caption{ The de-reddened line fluxes plotted in the Fig.~\ref{plot:line} in units of $\times10^\mathrm{-14}$ \,ergs\,cm$^\mathrm{-2}$\,s$^\mathrm{-1}$.}
    \label{tab:flux}

    \begin{tabular}{cccccccccccc} 
    \hline		
    \multicolumn{1}{c}{JD}&\multicolumn{10}{c}{Flux}\\
    
    \multicolumn{1}{c}{since 2400000}& 
    \multicolumn{1}{c}{H$\alpha$} & \multicolumn{1}{c}{Error}& 
    \multicolumn{1}{c}{H$\beta$} & \multicolumn{1}{c}{Error}& 
    \multicolumn{1}{c}{H$\gamma$} & \multicolumn{1}{c}{Error}& 
    \multicolumn{1}{c}{H$\delta$} & \multicolumn{1}{c}{Error}& 
    \multicolumn{1}{c}{He II 4686\,\AA} & \multicolumn{1}{c}{Error}& \\

    \hline

    59376.36 &   892.0 &        50.0 &  367.0 &       26.0 &   180.0 &        16.0 &   125.0 &        14.0 &      269.0 &           18.0 \\
    59410.25 &  1048.0 &        57.0 &  476.0 &       24.0 &   231.0 &        12.0 &   180.0 &        10.0 &      379.0 &           17.0 \\
    59477.11 &  1491.0 &        80.0 &  465.0 &       21.0 &   179.2 &         8.6 &   110.9 &         6.1 &      238.0 &           10.0 \\
    59648.46 &   741.0 &        43.0 &  165.5 &        8.8 &    59.6 &         3.9 &    33.6 &         2.7 &      103.5 &            5.9 \\
    59673.45 &   --    &        --   &  167.6 &        9.8 &    48.2 &         3.4 &    25.4 &         2.4 &      102.0 &            6.3 \\
    59714.43 &   595.0 &        39.0 &  176.3 &        9.5 &    66.5 &         4.1 &    43.8 &         2.9 &      105.3 &            6.2 \\
    59822.17 &   742.0 &        55.0 &  212.0 &       13.0 &    77.9 &         4.3 &    47.4 &         3.0 &      115.4 &            6.0 \\

    \hline	
    \end{tabular}

    \begin{tabular}{cccccccccccc} 
    \hline		
    \multicolumn{1}{c}{JD}&\multicolumn{10}{c}{Flux}\\
    
    \multicolumn{1}{c}{since 2400000}& 
    \multicolumn{1}{c}{He I 4471\,\AA} & \multicolumn{1}{c}{Error}&
    \multicolumn{1}{c}{He I 5876\,\AA} & \multicolumn{1}{c}{Error}&
    \multicolumn{1}{c}{He I 6678\,\AA} & \multicolumn{1}{c}{Error}& 
    \multicolumn{1}{c}{He I 7065\,\AA} & \multicolumn{1}{c}{Error}& 
    \multicolumn{1}{c}{He I 7280\,\AA} & \multicolumn{1}{c}{Error}&\\ 

    \hline
    
    59376.36 &      19.0 &           9.5 &      82.4 &          26.0 &      49.2 &           6.0 &      39.5 &           5.5 &     12.10 &          4.10 \\
    59410.25 &      49.4 &           5.8 &     111.0 &          36.0 &      65.6 &           4.7 &      53.8 &           4.3 &     18.70 &          2.70 \\
    59477.11 &      16.5 &           2.3 &      75.6 &          23.0 &      31.4 &           2.1 &      45.2 &           2.9 &      5.91 &          0.96 \\
    59648.46 &       4.1 &           1.0 &      30.2 &          10.0 &      21.0 &           2.2 &      19.8 &           2.5 &      4.50 &          1.60 \\
    59673.45 &       3.8 &           1.2 &      39.2 &          14.0 &      26.0 &           2.9 &      21.8 &           3.0 &      3.80 &          1.80 \\
    59714.43 &       6.1 &           1.3 &      30.9 &           9.4 &      22.3 &           2.1 &      18.1 &           2.2 &      4.40 &          1.50 \\
    59822.17 &       5.0 &           1.2 &      33.2 &           8.5 &      18.5 &           1.8 &      16.0 &           2.2 &      3.40 &          1.30 \\

    \hline	
    \end{tabular}
\end{table*}

\begin{table*}
    \centering
    \begin{tabular}{cccccccccccc} 
    \hline		
    \multicolumn{1}{c}{JD}&\multicolumn{10}{c}{Flux}\\

    \multicolumn{1}{c}{since 2400000}& 
    \multicolumn{1}{c}{Bowen 4640\,\AA} & \multicolumn{1}{c}{Error}& 
    \multicolumn{1}{c}{O [III] 5007 + He I 5016\,\AA} & \multicolumn{1}{c}{Error}& 
    \multicolumn{1}{c}{O VI 6825\,\AA} & \multicolumn{1}{c}{Error}& 
    \multicolumn{1}{c}{O I 7774\,\AA} & \multicolumn{1}{c}{Error}& 
    \multicolumn{1}{c}{O I 8444\,\AA} & \multicolumn{1}{c}{Error}& \\
    \hline

    59376.36 &       110.0 &            15.0 &       91.0 &           12.0 &      --   &           --  &     15.7 &          4.0 &     51.5 &          4.7 \\
    59410.25 &       112.8 &             8.6 &      108.0 &            7.6 &      13.6 &           4.1 &     18.8 &          2.8 &     67.4 &          3.5 \\
    59477.11 &        43.3 &             3.2 &       50.5 &            2.8 &      21.1 &           1.9 &      4.2 &          1.1 &     54.1 &          3.0 \\
    59648.46 &         8.1 &             1.6 &       19.4 &            1.8 &      19.5 &           2.6 &      4.2 &          2.0 &     45.4 &          4.3 \\
    59673.45 &        11.9 &             1.7 &       20.6 &            2.2 &      25.2 &           3.4 &      --  &          --  &     --   &          --  \\
    59714.43 &         7.9 &             1.9 &       24.7 &            2.2 &      15.9 &           2.2 &      3.2 &          1.5 &     30.5 &          2.8 \\
    59822.17 &        10.4 &             1.7 &       23.8 &            2.0 &      14.4 &           2.1 &      3.1 &          1.4 &     39.3 &          2.8 \\

    \hline	
    \end{tabular}
\end{table*}

\bsp	
\label{lastpage}
\end{document}